\documentclass[format=acmsmall, review=false, screen=true]{acmart}

\usepackage{colortbl}
\usepackage{cleveref}
\usepackage[utf8]{inputenc}
\crefname{section}{§}{§§}
\Crefname{section}{§}{§§}
\usepackage{enumitem}
\usepackage{fontawesome}
\usepackage{makecell}
\usepackage[10pt]{moresize}
\usepackage{pifont}

\usepackage{booktabs}
\usepackage{amsfonts}
\usepackage[font={sf,small}]{caption}
\usepackage[T1]{fontenc}
\usepackage{graphicx}
\usepackage{multirow}
\usepackage{soul}
\usepackage{textcomp}

\newcommand{\maciej}[1]{}

\newcommand{\noAnswer}{\textcolor{lightgray}{\faQuestionCircle}}
\newcommand{\commentline}[1]{& & & & & & & & & & & & & & & & & & & &#1}

\renewcommand{\faThumbsUp}{{\faBatteryFull}}
\renewcommand{\faThumbsOUp}{{\faBatteryHalf}}
\renewcommand{\faThumbsODown}{{\faTimes}}

\acmDOI{0000001.0000001}

\newif\iftr
\newif\ifconf
\newif\ifall

\trtrue
\conffalse
\allfalse

\newif\ifdel
\delfalse

\ifconf
\else
\fi

\begin{document}

\title[Survey and Taxonomy of Graph Databases]{Demystifying Graph Databases: Analysis and Taxonomy \\ of Data Organization, System Designs, and Graph Queries}

\author{Maciej Besta}
\affiliation{
  \institution{Department of Computer Science, ETH Zurich}
  \city{Zurich}
  \country{Switzerland}
}

\author{Robert Gerstenberger}
\affiliation{
  \institution{Department of Computer Science, ETH Zurich}
  \city{Zurich}
  \country{Switzerland}
}

\author{Emanuel Peter}
\affiliation{
  \institution{Department of Computer Science, ETH Zurich}
  \city{Zurich}
  \country{Switzerland}
}

\author{Marc Fischer}
\affiliation{
  \institution{PRODYNA (Schweiz) AG}
  \city{Zurich}
  \country{Switzerland}
}

\author{Michał Podstawski}
\affiliation{
  \institution{Future Processing}
  \city{Gliwice}
  \country{Poland}
}

\author{Claude Barthels}
\affiliation{
  \institution{Department of Computer Science, ETH Zurich}
  \city{Zurich}
  \country{Switzerland}
}

\author{Gustavo Alonso}
\affiliation{
  \institution{Department of Computer Science, ETH Zurich}
  \city{Zurich}
  \country{Switzerland}
}

\author{Torsten Hoefler}
\affiliation{
  \institution{Department of Computer Science, ETH Zurich}
  \city{Zurich}
  \country{Switzerland}
}

\begin{abstract}
\iftr
Graph processing has become an important part of multiple areas of computer
science, such as machine learning, computational sciences, medical
applications, social network analysis, and many others. Numerous graphs such as
web or social networks may contain up to trillions of edges. Often, these
graphs are also dynamic (their structure changes over time) and have
domain-specific rich data associated with vertices and edges.
Graph database systems such as Neo4j enable storing, processing, and analyzing
such large, evolving, and rich datasets. Due to the sheer size of such
datasets, combined with the irregular nature of graph processing, these systems
face unique design challenges.
To facilitate the understanding of this emerging domain, we present the first
survey and taxonomy of graph database systems. We focus on identifying and
analyzing fundamental categories of these systems (e.g., triple stores, tuple
stores, native graph database systems, or object-oriented systems), the
associated graph models (e.g., RDF or Labeled Property Graph), data
organization techniques (e.g., storing graph data in indexing structures or
dividing data into records), and different aspects of data distribution and
query execution (e.g., support for sharding and ACID). 51 graph
database systems are presented and compared, including Neo4j,
OrientDB, and Virtuoso.
%
%
We outline graph database queries and relationships with associated domains
(NoSQL stores, graph streaming, and dynamic graph algorithms).
%
%
Finally, we describe research and engineering challenges to outline the future
of graph databases.
\fi
\end{abstract}

\iftr

\ccsdesc[500]{General and reference~Surveys and overviews}
\ccsdesc[500]{Information systems~Data management systems}
\ccsdesc[500]{Information systems~Graph-based database models}
\ccsdesc[500]{Information systems~Data structures}
\ccsdesc[500]{Information systems~DBMS engine architectures}
\ccsdesc[500]{Information systems~Database query processing}
\ccsdesc[500]{Information systems~Parallel and distributed DBMSs}
\ccsdesc[300]{Information systems~Database design and models}
\ccsdesc[100]{Information systems~Distributed database transactions}
\ccsdesc[300]{Theory of computation~Data modeling}
\ccsdesc[300]{Theory of computation~Data structures and algorithms for data management}
\ccsdesc[100]{Theory of computation~Distributed algorithms}
\ccsdesc[100]{Computer systems organization~Distributed architectures}

\keywords{Graphs, Graph Databases, NoSQL Stores, Graph Database Management
Systems, Graph Models, Data Layout, Graph Queries, Graph Transactions, Graph
Representations, RDF, Labeled Property Graph, Triple Stores, Key-Value Stores,
RDBMS, Wide-Column Stores, Document Stores}

\fi

\maketitle

%
%
\renewcommand{\shortauthors}{M. Besta et al.}

\section{INTRODUCTION}

Graph processing is behind numerous problems in computing, for example in
medicine, machine learning, computational sciences, and
others~\cite{DBLP:journals/ppl/LumsdaineGHB07, jiang2011short}.
%
%
\iftr 
Graph algorithms are inherently difficult to design because of challenges such
as large sizes of processed graphs, little locality, or irregular
communication~\cite{DBLP:journals/ppl/LumsdaineGHB07, tate2014programming,
ching2015one, roy2015chaos, lin2018shentu, besta2019slim}. 
\fi
The difficulties are
increased by the fact that many such graphs are also dynamic (their structure
changes over time) and have rich data, for example arbitrary properties or labels, associated with vertices and edges.

\begin{figure}[h!]
\iftr
\vspace{-0.5em}
\fi
\centering
\includegraphics[width=0.97\textwidth]{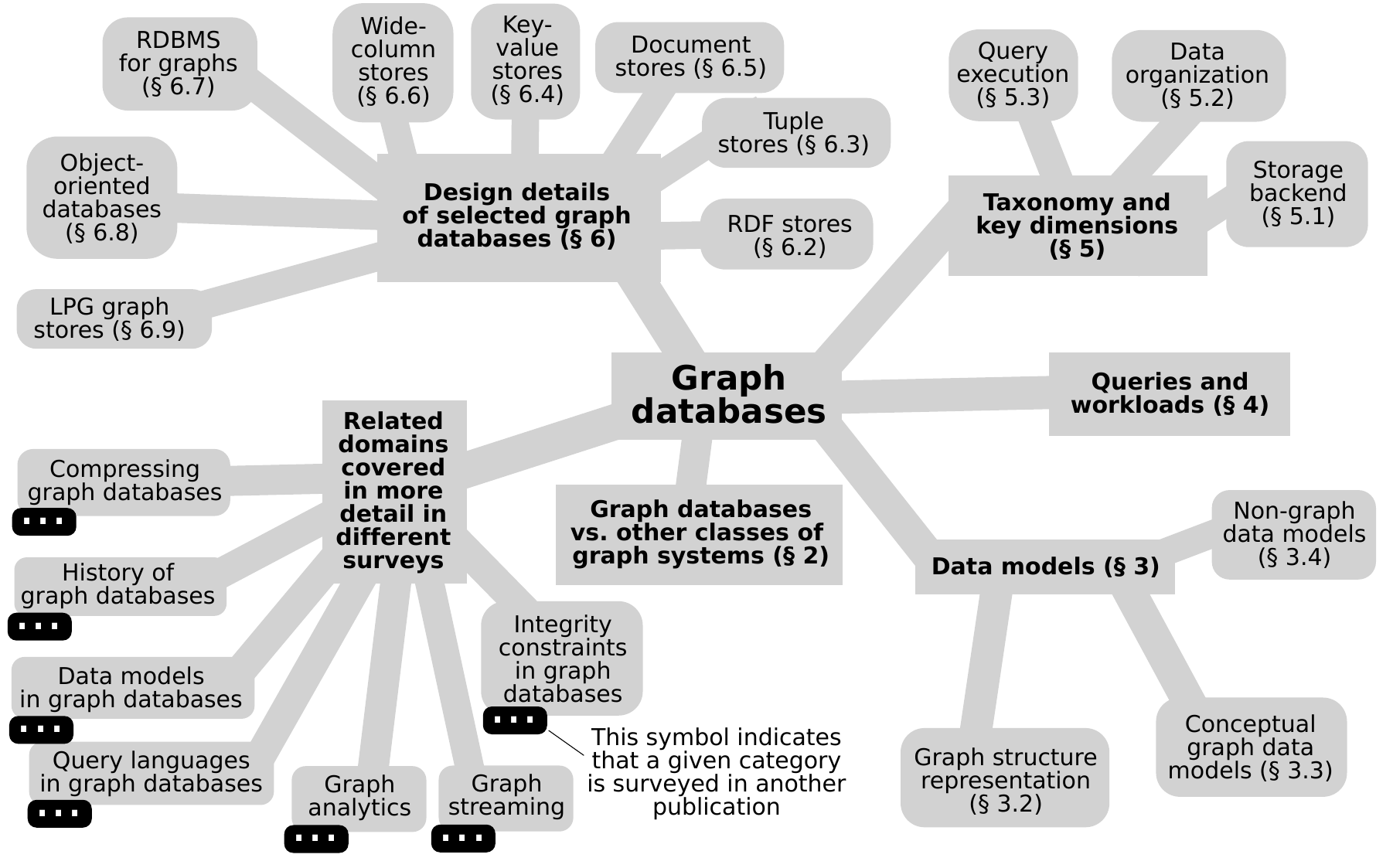}
\iftr
\vspace{-1em}
\fi
\caption{The illustration of the considered areas of graph databases.}
\label{fig:areas}
\iftr
\vspace{-1em}
\fi
\end{figure}

Graph databases (GDBs) such as
Neo4j~\cite{neo4j_book} emerged to enable storing, processing, and analyzing
large, evolving, and rich graph datasets.
%
%
Graph databases face unique challenges due to overall properties of irregular graph
computations combined with 
the demand for low latency and
high throughput of graph queries that can be both \emph{local} (i.e., accessing
or modifying a small part of the graph, for example a single edge) and
\emph{global} (i.e., accessing or modifying a large part of the graph, for
example all the edges). 
Many of these challenges belong to the following areas: ``general design''
(i.e., what is the most advantageous general structure of a graph database
engine), ``data models and organization'' (i.e., how to model and store the
underlying graph dataset), ``data distribution'' (i.e., whether and how to
distribute the data across multiple servers), and ``transactions and queries''
(i.e., how to query the underlying graph dataset to extract useful
information). 
This distinction is illustrated in Figure~\ref{fig:areas}.
In this work, we present the first survey and taxonomy on these system aspects
of graph databases.
%
%

In general, we provide the following contributions:


%
%
%

\begin{itemize}
\item We provide the first taxonomy of graph databases\footnote{Lists of graph databases can be found at\\
\url{http://nosql-database.org} \\ \url{https://database.guide} \\
\url{https://www.g2.com/categories/graph-databases} \\
\url{https://www.predictiveanalyticstoday.com/top-graph-databases} \\
\url{https://db-engines.com/en/ranking/graph+dbms}}, identifying and
analyzing key dimensions in the design of graph database systems.
\item We use our taxonomy to survey, categorize, and compare 51 graph database systems.
\item We discuss in detail the design of selected graph databases.
\item We outline related domains, such as queries and workloads in graph databases.
\item We discuss future challenges in the design 
of graph databases.
\end{itemize}

\subsection{Related Surveys}

There exist several surveys dedicated to the theory of graph databases.
In 2008, Angles et al.~\cite{gdb_survey_paper_Angles} described the history of
graph databases, and, in particular, the used {data models}, data structures,
query languages, and integrity constraints.
In 2017, Angles et al.~\cite{gdb_query_language_Angles} analyzed in more detail
query languages for graph databases, taking both an edge-labeled and a property
graph model into account and studying queries such as graph pattern matching
and navigational expressions.
In 2018, Angles and Gutierrez provided an
overview~\cite{angles2018introduction} of basic notions in the graph database
landscape. While being related to our work, it is largely orthogonal, focusing
on historical developments (which we exclude), details of many graph database
models and query languages (we only focus on the ones routinely supported by
existing graph database systems), and it only sketches at a very high level a
few selected graph database systems.
Our work, instead, focuses primarily on graph database systems and the
details of their design, and analyzes in depth all other aspects (graph data
models, query languages, queries) \emph{through the perspective of being
supported in these systems}.
Also in 2018, Bonifati et al.~\cite{bonifati2018querying} provided an in-depth
investigation into querying graphs, focusing on numerous aspects of query
specification and execution.
Moreover, there are surveys that focus on NoSQL
stores~\cite{davoudian2018survey, han2011survey, gajendran2012survey} and
the Resource Description Framework (RDF)~\cite{DBLP:journals/corr/Ozsu16}.
There is no survey dedicated to the systems aspects of graph databases, except
for several brief papers that cover small parts of the domain (brief
  descriptions of a few systems, concepts, or
  techniques~\cite{gdb_management_huge_unstr_data,
  gdb_survey_paper_Kaliyar, kumar2015domain, pokorny2015graph,
  junghanns2017management}, a survey of graph processing
  ubiquity~\cite{Sahu_gdb_online_survey}, and performance evaluations of
  a few systems~\cite{Vojtech_masterthesis_comp_gdb,
  vaikuntam2014evaluation, mccoll2014performance}).

\section{GRAPH DATABASES AND OTHER CLASSES OF GRAPH SYSTEMS}

Graph database systems are described in the literature as \emph{``systems
specifically designed for managing graph-like data following the basic
principles of database systems, i.e., persistent data storage, physical/logical
data independence, data integrity, and
consistency''}~\cite{angles2018introduction}.
However, other systems can also store and
process dynamic graphs. We now briefly discuss relations to three such
classes: other classes of databases, streaming graph frameworks, and general static graph processing
systems.

\subsection{{Graph Databases vs.~NoSQL Stores and Other Databases}}

NoSQL stores address various deficiencies of relational database systems, such
as little support for flexible data models~\cite{davoudian2018survey}. Graph
databases such as Neo4j can be seen as one particular type of NoSQL stores;
these systems are sometimes referred to as \emph{``native'' graph
databases}~\cite{neo4j_book}. Other types of NoSQL systems include
\emph{wide-column stores}, \emph{document stores}, and general \emph{key-value
stores}~\cite{davoudian2018survey}.
%
%
We focus on both ``native'' graph databases such as
Neo4j~\cite{neo4j_book} and on other systems used
specifically for maintaining graphs (relational databases, object-oriented
databases, NoSQL, and others).
\iftr
Figure~\ref{fig:considered_systems} shows the types of considered systems.
\begin{figure}[h]
\centering
\includegraphics[width=0.9\textwidth]{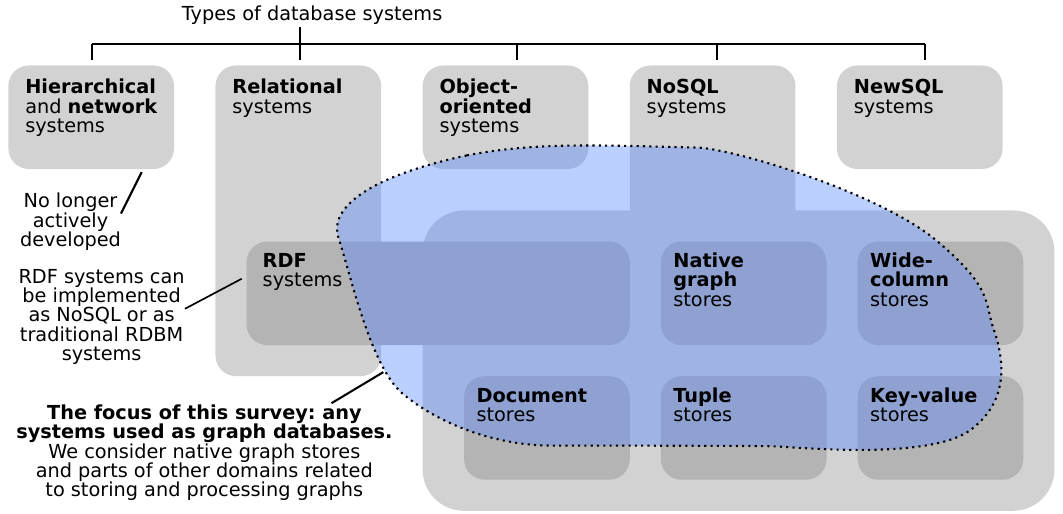}
\caption{The illustration of the considered types of databases.}
\label{fig:considered_systems}
\end{figure}
\fi

\subsection{{Graph Databases vs.~Graph Streaming Frameworks}}

In \emph{graph streaming}~\cite{besta2019practice}, the input graph is passed
as a stream of updates, allowing to add and remove edges in a simple way. Graph
databases are related to graph streaming in that they face graph updates of
various types. Still, they usually deal with complex graph models (such as the
Labeled Property Graph (LPG)~\cite{gdb_query_language_Angles} or Resource Description
Framework~\cite{rdf_links}) where both vertices and edges may be of different
types and may be associated with arbitrary properties.  Contrarily, graph
streaming frameworks focus on simple
graph models where edges or vertices may have weights and, in some cases,
simple additional properties such as time stamps. Moreover, challenges in the
design of graph databases include transactional support, persistence,
physical/logical data independence, data integrity, or consistency; these topics are
little related to graph streaming frameworks.

\subsection{{Graph Databases vs.~Static Graph Processing Systems}}
\label{sec:static-g-s}

A lot of effort has been dedicated to static graph analytics~\cite{han2014experimental,
doekemeijer2014survey, shi2018graph, batarfi2015large, mccune2015thinking,
yan2016big}.
\iftr
Many of these works focus on the vertex-centric paradigms~\cite{besta2019graph,
ahn2016scalable, kalavri2017high, shi2018graph}. Some works also focus on
edge-centric or linear algebra paradigms~\cite{song2018graphr,
sundaram2015graphmat, kepner2016mathematical}.
\fi
The key differences to graph databases are that graph processing systems
usually focus on graphs that are static and simple, i.e., do not have rich
attached data such as labels or key-value pairs (details
in~\cref{sec:g_models}). 
Moreover, static graph processing
systems do not focus on topics such as transactions, persistence,
physical/logical data independence, data integrity, or consistency.
%

\vspace{0.25em}
\noindent
Graph streaming frameworks
and static graph processing systems are \emph{not} covered in this work.

\section{GRAPH DATA MODELS IN THE LANDSCAPE OF GRAPH DATABASES}
\label{graph_models_section}

We start with data models. This includes conceptual graph models and
representations, and non-graph models used in graph databases.
Key symbols and abbreviations are shown in Table~\ref{tab:symbols}.

\begin{table}[t]
\centering
\iftr
\small
\fi
\sf
\begin{tabular}{ll@{}}
\toprule
                    $G$ & A graph $G=(V,E)$ where $V$ is a set of vertices and $E$ is a set of edges.\\
                    $n,m$&The count of vertices and edges in a graph $G$; $|V| = n, |E| = m$.\\
                    $d, \hat{d}$&The average degree and the maximum degree in a given graph, respectively.\\
                    $\mathcal{P}(S) = 2^S$ & The power set of $S$: a set that contains all possible subsets of $S$.\\
%
%
AM, $\mathbf{M}$ & The Adjacency Matrix representation. $\mathbf{M} \in \{0,1\}^{n,n}$, $\mathbf{M}_{u,v}=1 \Leftrightarrow (u,v)\in E$.\\
AL, $\mathnormal{A}_{u}$ & The Adjacency List representation and the adjacency list of a vertex~$u$; $v\in \mathnormal{A}_{u} \Leftrightarrow (u,v)\in E$.\\
%
%
LPG, RDF & Labeled Property Graph (\cref{lpg_defininion}) and Resource Description Framework (\cref{rdf_section}). \\
KV, RDBMS & Key-Value store (\cref{keyvalue_section}) and Relational Database Management Systems (\cref{rdbms_section}). \\
OODBMS & Object-Oriented Database Management Systems (\cref{objectoriented_section}). \\
OLTP, OLAP & Online Transaction Processing (\cref{typesoftransactions_section}) and Online Analytics Processing (\cref{typesoftransactions_section}). \\
ACID & Transaction guarantees (Atomicity, Consistency, Isolation, Durability). \\
\bottomrule
\end{tabular}
\caption{The most relevant symbols and abbreviations used in this work.}
\label{tab:symbols}
\end{table}

\subsection{Simple Graph Model}
\label{simplegraph_section}

We start with a simple graph model that is a basis for more complex and
richer conceptual graph models used in graph databases.
A graph $G$ can be modeled as a tuple $(V,E)$ where $V$ is a set of vertices
and $E \subseteq V \times V$ is a set of edges. $G = (V,E)$ can also be denoted
as $G(V,E)$. We have $|V|=n$ and $|E|=m$. For a directed $G$, an edge
$e=(u,v)\in E$ is a tuple of two vertices, where $u$ is the out-vertex (also
called ``source'') and $v$ is the in-vertex (also called ``destination''). If
$G$ is undirected, an edge $e=\{u,v\}\in E$ is a set of two vertices. Finally,
a weighted graph~$G$ is modeled with a triple $(V, E, w)$; $w: E \to
\mathbb{R}$ maps edges to weights.

\subsection{Fundamental Representations of Graph Structure}
\label{sec:g_reps}

We also summarize two fundamental ways to represent the structure of
connections between vertices.
Two common such graph representations of vertex neighborhoods are
the \emph{adjacency matrix} format (AM) and the \emph{adjacency list} format
(AL). 
\iftr
We illustrate these representations in Figure~\ref{fig:reps}.
\fi

\iftr
\begin{figure}[h!]
\centering
\includegraphics[width=0.8\textwidth]{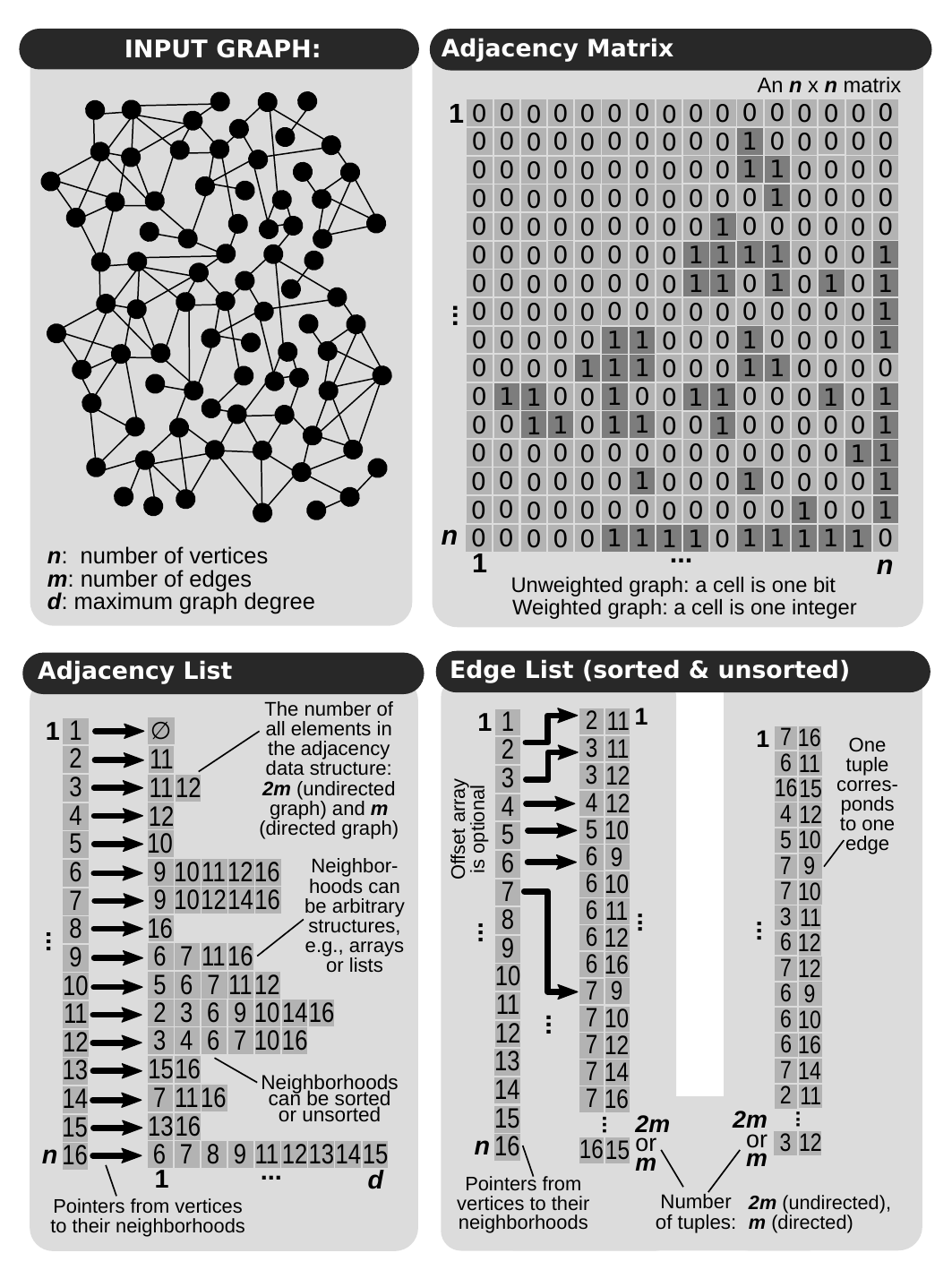}
\caption{\textbf{Illustration of fundamental graph representations:} 
Adjacency Matrix, Adjacency List, and Edge List.
}
\label{fig:reps}
\end{figure}
\fi

In the AM format, a matrix $\mathbf{M} \in \{0,1\}^{n,n}$ determines the
connectivity of vertices:
$
\mathbf{M}_{u,v}=1 \Leftrightarrow (u,v)\in E.
$
In the AL format, each vertex $u$ has an associated adjacency list
$\mathnormal{A}_{u}$. This adjacency list maintains the IDs of all
vertices adjacent to $u$. 
%
%
Each adjacency list is often stored as a contiguous array of vertex IDs.
We have
$
v\in \mathnormal{A}_{u} \Leftrightarrow (u,v)\in E.
$
%
%

AM uses $\mathcal{O}\left(n^2\right)$ space and can check connectivity of two
vertices in $\mathcal{O}\left(1\right)$ time. AL requires $\mathcal{O}\left(n
+ m\right)$ space and it can check connectivity in
$\mathcal{O} \left(|\mathnormal{A}_u| \right) \subseteq
\mathcal{O} \left(\hat{d} \right)$ time.
The AL or AM representations
are used to maintain the graph structure (i.e., neighborhoods of vertices).

\subsection{Conceptual Graph Data Models Used in Graph Databases}
\label{sec:g_models}

We now introduce the conceptual graph models used by the surveyed systems;
these models extend the simple graph model from~\cref{simplegraph_section}.
A simple graph model is often used in graph frameworks such as
Pregel~\cite{Malewicz:2010:PSL:1807167.1807184} or
STINGER~\cite{ediger2012stinger}. However, it is not commonly used with graph
databases.

\subsubsection{Hypergraph Model}
\label{sec:hg_model}

A hypergraph $H$ generalizes a simple graph: any of its edges can join
\emph{any number of vertices}.  Formally, a
hypergraph is also modeled as a tuple $(V,E)$ with $V$ being a set of vertices. $E$ is defined as $E
\subseteq (\mathcal{P}(V) \setminus \emptyset)$ and it contains \emph{hyperedges}:
non-empty subsets of $V$.

Hypergraphs are rarely used in graph databases and graph processing systems.
In this survey, we describe a system called HyperGraphDB (\cref{hyperGDB_section})
that focuses on storing and querying hypergraphs. 

\subsubsection{Labeled Property Graph Model}
\label{lpg_defininion}

\begin{figure*}[t]
\centering
\includegraphics[width=0.6\textwidth]{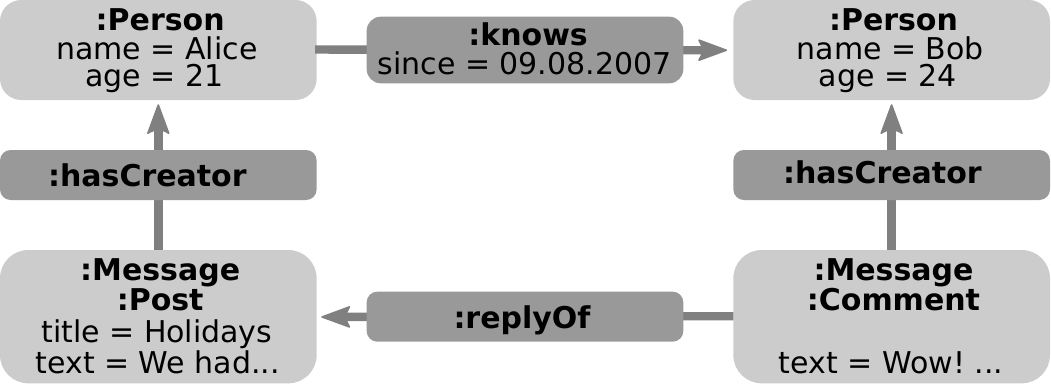}
\caption{
\textbf{The illustration of an example Labeled Property Graph (LPG).} Vertices
and edges can have labels (bold, prefixed with colon) and properties (key =
value).  We present a subgraph of a social network, where a person can know
other persons, post messages, and comment on others' messages.
}
\label{fig:lpg}
\end{figure*}

The classical graph model, a tuple $G = (V,E)$, is adequate for many problems
such as computing vertex centralities~\cite{brandes2001faster}.  However, it is
not rich enough to model various real-world problems. This is why graph
databases often use the \emph{Labeled Property Graph Model}, sometimes
simply called a property graph~\cite{bonifati2018querying, gdb_query_language_Angles}. In LPG, one
augments the simple graph model $(V,E)$ with \textit{labels} that define
different subsets (or classes) of vertices and edges. Furthermore, every vertex
and edge can have any number of \textit{properties}~\cite{bonifati2018querying} (often also called
\textit{attributes}). A property is a pair $(key, value)$, where \textit{key}
identifies a property and \textit{value} is the corresponding value of this
property~\cite{bonifati2018querying}.
Formally, an LPG is defined as a tuple 
$
(V,E,L,l_V,l_E,K,W,p_V,p_E)
$
where $L$ is the set of labels.  $l_V:V\mapsto \mathcal{P}(L)$ and $l_E:E\mapsto
\mathcal{P}(L)$ are labeling functions. Note that $\mathcal{P}(L)$ is the power
set of $L$, denoting all the possible subsets of $L$. Thus, each vertex and edge
is mapped to a subset of labels. Next, a vertex as well as an edge can be associated with 
any number of properties. We model a property as a key-value pair
$p=(key, value)$, where $key \in \mathnormal{K}$ and $value \in \mathnormal{W}$.
$\mathnormal{K}$ and $\mathnormal{W}$ are sets of all possible keys and values.
Finally,
$p_V(u)$ denotes the set of property key-value pairs of the vertex $u$,
$p_E(e)$ denotes the set of property key-value pairs of the edge $e$.
An example LPG is in Figure~\ref{fig:lpg}.
Note that, in LPGs, $E$ may be a multi-set (i.e., there may be more than a
single edge between vertices, even having identical labels and/or key-value
sets.
All systems considered in this work use some variant of the LPG,
with the exception of RDF systems or when explicitly discussed.

\subsubsection{Variants of Labeled Property Graph Model}
\label{lpgvariants_section}

Several databases support variants of LPG.
First, Neo4j~\cite{neo4j_book} (a graph database described in detail
in~\cref{neo4j_design}) supports an arbitrary number of labels for vertices.
However it only allows for one label, (called \emph{edge-type}), per edge.
Next, ArangoDB~\cite{arangodb_links} (a graph database described in detail
in~\cref{arangodb_design}) only allows for one label per vertex (\emph{vertex-type})
and one label per edge (\emph{edge-type}).  This facilitates the separation of
vertices and edges into different document collections.
Moreover, edge-labeled graphs~\cite{gdb_query_language_Angles} do not allow for
any properties and use labels in a restricted way. Specifically, only edges
have labels and each edge has exactly one label. Formally, $G=(V,E,L)$, where
$V$ is the set of vertices and $E \subseteq V\times L \times V$ is the set of
edges. Note that this definition enables two vertices to be connected by
multiple edges with different labels.
Finally, some effort was dedicated to LPG variants that facilitate storing
historical graph data~\cite{castelltort2013representing}.

\subsubsection{Resource Description Framework}
\label{rdf_section}

The \emph{Resource Description Framework}~\cite{rdf_links} is a
collection of specifications for representing information. It was introduced by
the World Wide Web Consortium (W3C) in 1999 and the latest version (1.1) of the
RDF specification was published in 2014. Its goal is to enable a simple format
that allows for easy data exchange between different formats of data. It is
especially useful as a description of irregularly connected data. The core part
of the RDF model is a collection of \emph{triples}. Each triple consists of a
\emph{subject}, a \emph{predicate}, and an \emph{object}. Thus, RDF databases
are also often called \emph{triple stores} (or \emph{triplestores}).  Subjects
can either be identifiers (called Uniform Resource Identifiers (URIs)) or blank
nodes (which are dummy identifiers for internal use). Objects can be URIs,
blank nodes, or literals (which are simple values). With triples, one can
connect identifiers with identifiers or identifiers with literals. The
connections are named with another URI (the predicate).
RDF triples can be formally described as
$$ (s, p, o) \in (URI \cup blank) \times (URI) \times (URI \cup blank \cup
literal) $$
$s$ represents a subject, $p$ models a predicate, and $o$ represents an object.
$URI$ is a set of Uniform Resource Identifiers; $blank$ is a set of blank node
identifiers, that substitute internally URIs to allow for more complex data
structures; $literal$ is a set of literal values~\cite{rdfmodel,
DBLP:journals/corr/Ozsu16}.

\subsubsection{Transformations between LPG and RDF}
\label{lpg_rdf_transform_section}

\begin{figure}[h!]
\centering
\includegraphics[width=0.85\textwidth]{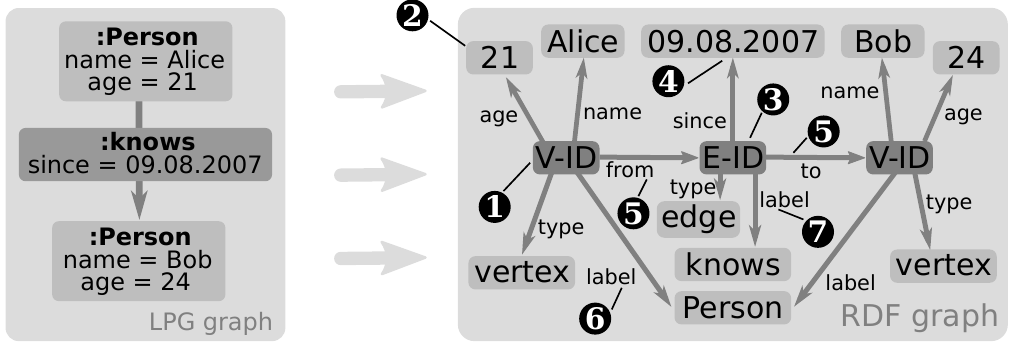}
\caption{\textbf{Comparison of an LPG and an RDF graph:} a transformation from
LPG to RDF.  ``V-ID'', ``E-ID'', ``age'', ``name'', ``type'', ``from'', ``to'',
``since'' and ``label'' are RDF URIs.  Numbers in black circles refer to
transformation steps in~\cref{lpg_rdf_transform_section}.}
\label{fig:rdf:transformation_1}
%
%
\end{figure}

To represent a Labeled Property Graph in the RDF model, LPG vertices are mapped
to URIs (\ding{182}) and then RDF triples are used to link those vertices with
their LPG properties by representing a property key and a property value with,
respectively, an RDF predicate and an RDF object (\ding{183}). For example, for
a vertex with an ID \textit{vertex-id} and a corresponding property with a key
\textit{property-key} and a value \textit{property-value}, one creates an RDF
triple \textit{(vertex-id, property-key, property-value)}. Similarly, one can
represent edges from the LPG graph model in the RDF model by giving each edge
the URI status (\ding{184}), and by linking edge properties with specific edges
analogously to vertices: \textit{(edge-id, property-key, property-value)}
(\ding{185}). Then, one has to use two triples to connect each edge to any of
its adjacent vertices (\ding{186}). Finally, LPG labels can also be transformed
into RDF triples in a way similar to that of properties~\cite{neo4j_RDF_LPG},
by creating RDF triples for vertices (\ding{187}) and edges (\ding{188}) such
that the predicate becomes a ``label'' URI and contains the string name of this
label.
Figure~\ref{fig:rdf:transformation_1} shows an example of transforming an LPG
graph into RDF triples.
More details on transformations between LPG and RDF are provided by Hartig~\cite{hartig2019foundations}.

\begin{figure}[h!]
\centering
\includegraphics[width=1.0\textwidth]{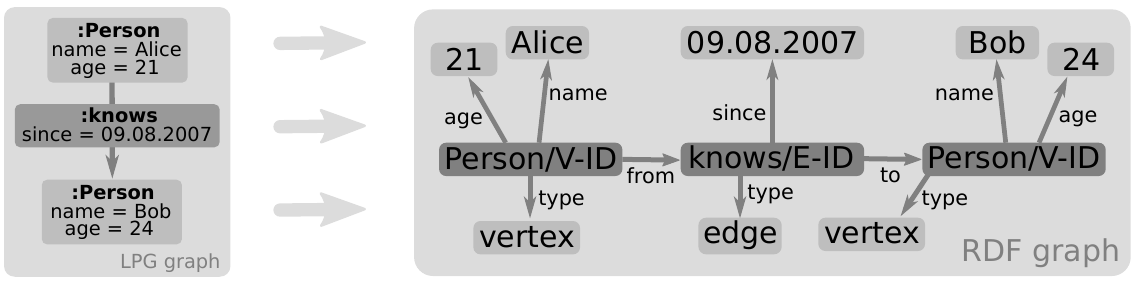}
\caption{\textbf{Comparison of an LPG and an RDF graph:} a 
transformation from LPG to RDF, given vertices and edges have only one
label. ``Person/V-ID'', ``knows/E-ID'', ``age'', ``name'', ``type'', ``from'',
``to'' and ``since'' are RDF URIs.}
\label{fig:rdf:transformation_2}
%
%
\end{figure}

If all vertices and edges only have one label, one can omit the triples for
labels and store the label (e.g., ``Person'') \emph{together with} the vertex
or the edge name (``V-ID'' and ``E-ID'') in the identifier. We illustrate a
corresponding example in Figure~\ref{fig:rdf:transformation_2}.

Transforming RDF data into the LPG model is more complex, since RDF predicates,
which would normally be translated into edges, are URIs. Thus, while deriving
an LPG graph from an RDF graph, one must map edges to vertices and link such
vertices, otherwise the resulting LPG graph may be disconnected.
There are several schemes for such an RDF to LPG transformation, for example
deriving an LPG graph which is bipartite, at the cost of an increased graph
size~\cite{rdfmodel}. Details and examples are provided in a report by
Hayes~\cite{rdfmodel}.


%

%

\subsection{Non-Graph Data Models and Storage Schemes Used in Graph Databases}

In addition to the conceptual graph models, graph databases also often
incorporate different storage schemes and data models that do not target
specifically graphs but are used in various systems to model and store graphs. 
These models include \emph{collections of key-value pairs}, \emph{documents},
and \emph{tuples} (used in different types of NoSQL stores), \emph{relations
and tables} (used in traditional relational databases), and \emph{objects}
(used in object-oriented databases).
Different details of these models and the database systems based on them are
described in other surveys, for example in a recent publication on NoSQL stores
by Davoudian et al.~\cite{davoudian2018survey}. Thus, we omit extensive
discussions and instead offer brief summaries, \emph{focusing on how they are
used to model or represent graphs}.

\subsubsection{Collection of Key-Value Pairs}
\label{kv_section}

Key-value stores are the simplest NoSQL stores~\cite{davoudian2018survey}.
Here, the data is stored as a collection of \textit{key-value} pairs, with the
focus on high-performance and highly-scalable lookups based on keys. The exact
form of both keys and values depends on a specific system or an application.
Keys can be simple (e.g., an URI or a hash) or structured.  Values are often
encoded as byte arrays (i.e., the structure of values is usually schema-less).
However, a key-value store can also impose some additional data layout,
structuring the schema-less values~\cite{davoudian2018survey}.

Due to the general nature of key-value stores, there can be many ways of
representing a graph as a collection of KV values. We describe several concrete
example systems~\cite{hypergraphdb_links, trinity_paper, dgraph_links, redisgraph_links} in~\cref{keyvalue_section}. For example, one can use vertex
labels as keys and encode the neighborhoods of vertices as values.

\subsubsection{Collection of Documents}
\label{doc_section}

A document is a fundamental storage unit in a class of NoSQL databases called
document stores~\cite{davoudian2018survey}. These documents are stored in
collections. Multiple collections of documents constitute a database.
A document is encoded using a selected standard semi-structured format, e.g., JSON~\cite{json_links}
or XML~\cite{xml_link}. Document stores extend key-value
stores in that a document can be seen as a value that has a certain flexible
\emph{schema}. This schema consists of \emph{attributes}, where each attribute
has a \emph{name} along with one or more \emph{values}.
Such a structure based on documents with attributes allows for various value
types, key-value pair storage, and recursive data storage (attribute values can
be lists or key-value dictionaries).

In all surveyed document stores~\cite{arangodb_links, orientdb_links,
azure_cosmosdb_links, bitsy_links, faunadb_links} (\cref{document_store_section}), each vertex is stored in a
vertex document. The capability of documents to store key-value pairs is used
to store vertex labels
and properties within the corresponding vertex document.  The details of edge
storage, however, are system-dependent: edges can be stored in the document
corresponding to the source vertex of each edge, or in the documents of the
destination vertices.  As documents do not impose any restriction on what
key-value pairs can be stored, vertices and edges may have different sets of
properties.

\subsubsection{Collection of Tuples}
\label{tuple_model_section}

Tuples are a basis of NoSQL stores called tuple stores. A tuple store
generalizes an RDF store: RDF stores are restricted to triples (or -- in some cases -- 4-tuples, also referred to as \emph{quads}) whereas
tuple stores can contain \emph{tuples of an arbitrary size}. Thus, the number
of elements in a tuple is not fixed and can vary, even within a single
database. Each tuple has an ID which may also be a direct memory pointer.

A collection of tuples can model a graph in different ways. For example, one
tuple of size $n$ can store pointers to other tuples that contain neighborhoods
of vertices. The exact mapping between such tuples and graph data is specific
to different databases; we describe an example~\cite{whitedb_links} in~\cref{tuple_section}.

\subsubsection{Collection of Tables}
\label{table_section}

Tables are the basis of Relational Database Management Systems
(RDBMS)~\cite{hellerstein2005readings, codd1989relational,
atzeni1993relational}. Tables consist of rows and columns. Each row represents
a single data element, for example a car. A single column usually defines a
certain data attribute, for example the color of a car. Some columns can
define unique IDs of data elements, called \emph{primary keys}.
Primary keys can be used to implement relations between data elements.  A
one-to-one or a one-to-many relation can be implemented with a single
additional column that contains the copy of a primary key of the related data
element (such primary key copy is called the \emph{foreign key}).  A
many-to-many relation can be implemented with a dedicated table containing
foreign keys of related data elements.

To model a graph as a collection of tables, one can implement vertices and
edges as rows in two separate tables. Each vertex has a unique primary key that
constitutes its ID. Edges can relate to their source or destination vertices by
referring to their primary keys (as foreign keys). LPG labels and properties,
as well as RDF predicates, can be modeled with additional
columns~\cite{rdbms_allinone_paper, graphgen_rdbms_paper}.
We present and analyze different graph database systems~\cite{titan_link, oracle_spatial} based on tables
in~\cref{widecolumn_section} and~\cref{rdbms_section}.

\subsubsection{Collection of Objects}
\label{object_section}

One can also use collections of objects in Object-Oriented Database
Management Systems (OODBMS)~\cite{oodbms_paper} to model graphs. Here, data
elements and their relations are implemented as objects linked with some form
of pointers. The details of modeling graphs as objects heavily depend on
specific designs. We provide details for an example system~\cite{velocitygraph_links}
in~\cref{objectoriented_section}.

\section{QUERIES \& WORKLOADS IN THE LANDSCAPE OF GRAPH DATABASES}
\label{sec:queries-workloads}

We describe graph database workloads.

\subsection{OLAP and OLTP}
\label{typesoftransactions_section}

First, one distinguishes between \emph{Online Transactional Processing
(OLTP)} and \emph{Online Analytical Processing (OLAP)}.
OLTP queries are small, interactive, transactional, and local in
scope (i.e., they process only a small part of the graph). Examples are
neighborhood queries, lookups, inserts, deletes, and updates of single (or a
few) vertices and edges. 
OLAP queries are usually not processed at interactive speeds, as they are
inherently complex and global in scope (i.e., they span the whole graphs).
Examples are PageRank~\cite{pagerank_article} or Breadth-First Search (BFS).

Now, static graph processing systems (\cref{sec:static-g-s}) focus on OLAP. However,
many graph databases also support a rich set of
OLAP workloads. This includes Neo4j~\cite{neo4j_book}, Cray Graph
Engine~\cite{cge_paper}, Amazon
Neptune~\cite{amazon_neptune_links},
TigerGraph~\cite{tiger_graph_links}, and many others (see
Tables~\ref{survey-table1}-\ref{survey-table2}).  For example,
Neo4j provides algorithms for vertex centrality (e.g., PageRank, Betweenness
Centrality, Eigenvector Centrality), community detection (e.g., Louvain,
Triangle Counting, Weakly Connected Components, Label Propagation), graph
traversals (BFS, DFS), shortest paths (e.g., Delta Stepping, A$^*$), and many
others.
\emph{Thus, we focus on {both OLTP and OLAP, in the
context of how they are supported by graph databases.}}
%
%

\subsection{Graph Queries Beyond OLAP vs.~OLTP}

\begin{figure}[h]
\centering
\includegraphics[width=1.0\textwidth]{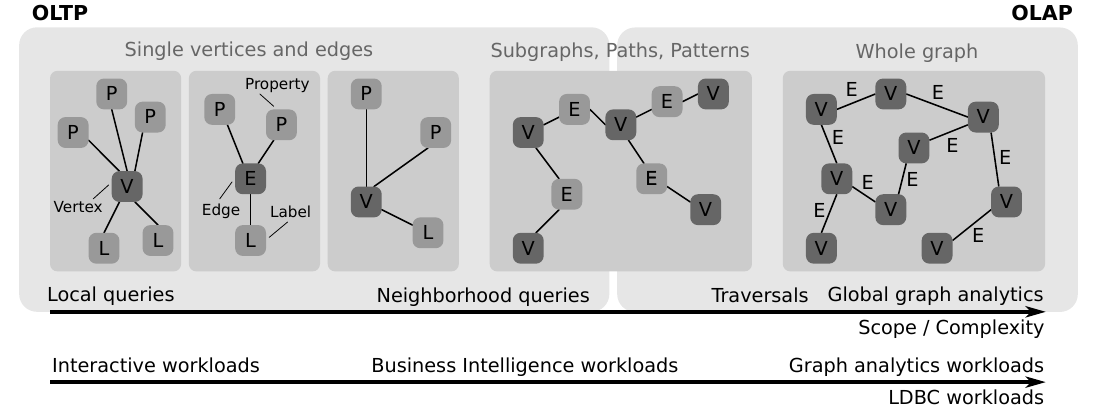}
\caption{Illustration of \textbf{different query scopes} and their relation
to other graph query taxonomy aspects, in the context of accessing a Labeled
Property Graph.}
\label{fig:lpg:hierarchy}
\end{figure}

We also offer an analysis of graph queries beyond the simple
distinction into OLAP and OLTP classes.
%
%
%
Figure~\ref{fig:lpg:hierarchy} illustrates the queries in the
context of accessing the LPG graph.
We omit detailed discussions and examples as they are provided in different
associated papers (query languages~\cite{gdb_query_language_Angles,
angles2018g}, analytics workloads~\cite{apache_giraph}, benchmarks related
to certain aspects~\cite{ciglan2012benchmarking, lissandrini2017evaluation,
lissandrini2018beyond} and whole systems~\cite{barahmand2013bg,
armstrong2013linkbench, ldbc, ldbc_snb_specification, ldbc_graphanalytics_paper,
early_ldbc_paper, capotua2015graphalytics,
jouili2013empirical} and surveys on system performance~\cite{dominguez2010survey,
mccoll2014performance}).
\iftr
Instead,
our goal is to deliver a broad overview, and point the reader to
the detailed material available elsewhere.
\fi

\subsubsection{Scopes of Graph Queries}
\label{sec:query_scope}

We describe queries in the increasing order of their scope. We focus on
the LPG model, see~\cref{lpg_defininion}. Figure~\ref{fig:lpg:hierarchy}
depicts the scope of graph queries.

\textbf{Local Queries}
Local queries involve single vertices or edges.  For example, given a
vertex or an edge ID, one may want to retrieve the labels and properties of
this vertex or edge. Other examples include
verifying whether a given vertex or a given edge have a given label (given the label
name), or whether they have a property of a specified type.
%
%
These queries are used in social network workloads~\cite{barahmand2013bg,
armstrong2013linkbench} (e.g., to fetch the profile information of a user) and
in benchmarks~\cite{jouili2013empirical} (e.g., to measure the vertex look-up
time).

\textbf{Neighborhood Queries}
Neighborhood queries retrieve all edges adjacent to a given vertex, or the
vertices adjacent to a given edge. This query can be further restricted by,
for example, retrieving only the edges with a specific label.
%
%
Similarly to local queries, social networks often require a retrieval of the
friends of a given person, which results in querying the local
neighborhood~\cite{barahmand2013bg, armstrong2013linkbench}.

\textbf{Traversals}
In a traversal query, one explores a part of the graph beyond a single
neighborhood. These queries usually start at a single vertex (or a small set of
vertices) and traverse some graph part. We call the initial vertex or the set
of vertices the \textit{anchor} or \emph{root} of the traversal. Queries can
restrict what edges or vertices can be retrieved or traversed. As this is a
common graph database task, this query is also used in different performance
benchmarks~\cite{ciglan2012benchmarking, jouili2013empirical,
dominguez2010survey}.

\textbf{Global Graph Analytics}
Finally, we identify global graph analytics queries, which
by definition consider the whole graph (not necessarily every property but all
vertices and edges). Different benchmarks~\cite{mccoll2014performance,
capotua2015graphalytics, dominguez2010survey, beamer2015gap} take these large-scale queries
into account since they are used in different fields such as threat
detection~\cite{eberle2010insider} or computational
chemistry~\cite{balaban1985applications}. As indicated in
Tables~\ref{survey-table1} and~\ref{survey-table2}, many graph databases
support such queries.
\iftr
Graph processing systems such as
Pregel~\cite{Malewicz:2010:PSL:1807167.1807184}
or Giraph~\cite{apache_giraph} focus
specifically on resolving global analytics~\cite{han2014experimental}. Example queries include resolving global
pattern matching~\cite{patternmatching_paper,patternmatching_paper2}, shortest
paths~\cite{Dijkstra1959}, max-flow or min-cut~\cite{mincutmaxflow_paper},
minimum spanning trees~\cite{sst_kruskal_paper}, diameter, eccentricity,
connected components, PageRank~\cite{pagerank_article}, and many others.
Some traversals can also be global (e.g., finding all shortest paths of unrestricted 
length), thus falling into the category of global analytics queries.
Global analytics workloads have been a subject of numerous research efforts
in the last decade~\cite{batarfi2015large, mccune2015thinking, kalavri2017high,
besta2017slimsell, besta2017push, solomonik2017scaling, plattner2009common}
\fi

\subsubsection{Classes of Graph Workloads}
\label{sec:workloads_ldbc}

We also outline an existing taxonomy of graph database workloads that is
provided as a part of the LDBC benchmarks~\cite{ldbc}. LDBC is an effort by
academia and industry to establish a set of standard benchmarks for measuring
the performance of graph databases. The effort currently specifies
\emph{interactive workloads}, \emph{Business Intelligence workloads}, and
\emph{graph analytics workloads}.
%

\textbf{Interactive Workloads}
A part of LDBC called the Social Network Benchmark
(SNB)~\cite{ldbc_snb_specification} identifies and analyzes \textit{interactive
workloads} that can collectively be described as either read-only queries or
simple transactional updates.  They are divided into three further categories.
First, \textit{short read-only queries} start with a single graph element
(e.g., a vertex) and lookup its neighbors or conduct small traversals. Second,
\textit{complex read-only queries} traverse larger parts of the graph; they are
used in the LDBC benchmark to not just assess the efficiency of the data
retrieval process but also the quality of query optimizers.
%
%
Finally, \textit{transactional update queries} insert, modify, remove either a single element
(e.g., a vertex), possibly together with its adjacent edges, or a single edge.
This workload tests common graph database operations such as the lookup of a
friend profile in a social network, or friendship removal.

\textbf{Business Intelligence Workloads}
Next, LDBC identifies \textit{Business Intelligence (BI)
workloads}~\cite{early_ldbc_paper}, which fetch large data volumes, spanning
large parts of a graph. Contrarily to the interactive workloads, the BI
workloads heavily use summarization and aggregation operations such as sorting,
counting, or deriving minimum, maximum, and average values. They are
read-only. The LDBC specification provides an extensive list of BI workloads
that were selected so that different performance aspects of a database
are properly stressed when benchmarking.

\textbf{Graph Analytics Workloads}
Finally, the LDBC effort comes with a graph analytics
benchmark~\cite{ldbc_graphanalytics_paper}, where six graph algorithms are
proposed as a standard benchmark for a graph analytics part of a graph
database. These algorithms are \emph{Breadth-First Search,
PageRank, weakly connected
components~\cite{gianinazzi2018communication}, community detection using label
propagation~\cite{boldi2011layered}, deriving the local clustering
coefficient~\cite{schaeffer2007graph}, and computing single-source shortest
paths~\cite{Dijkstra1959}}.

\textbf{LDBC Workloads}
The LDBC \textit{interactive workloads} correspond to \textit{local},
\textit{neighborhood}, and \textit{traversals}.  The LDBC \textit{Business
Intelligence workloads} range from \textit{traversals} to \textit{global graph
analytics queries}. The LDBC graph analytics benchmark corresponds to
\textit{global graph analytics}.

\subsubsection{Graph Patterns and Navigational Expressions}
\label{sec:graph_patterns}

Angles et al.~\cite{gdb_query_language_Angles} inspected in detail the theory
of graph queries.  In one identified family of graph queries, called
\emph{simple graph pattern matching}, one prescribes a graph pattern (e.g., a
specification of a class of subgraphs) that is then matched to the graph
maintained by the database, searching for the occurrences of this pattern.
This query can be extended with aggregation and a projection function to so
called \textit{complex graph pattern matching}. Next, \textit{path
queries} allow to search for paths of arbitrary distances in the graph.  One
can also combine complex graph pattern matching and path queries, resulting in
\textit{navigational graph pattern matching}, in which a graph pattern can be
applied recursively on the parts of the path.

\iftr

\subsubsection{Input Loading}
\label{sec:workloads_loading}

Finally, certain benchmarks also analyze bulk input
loading~\cite{jouili2013empirical, ciglan2012benchmarking,
dominguez2010survey}.  Specifically, given an input dataset, they measure the
time to load this dataset into a database.  This scenario is common when data
is migrated between systems.


\fi

\section{TAXONOMY OF GRAPH DATABASE SYSTEMS}
\label{sec:taxo}

\begin{figure*}[htbp]
\centering
\includegraphics[width=1.03\textwidth]{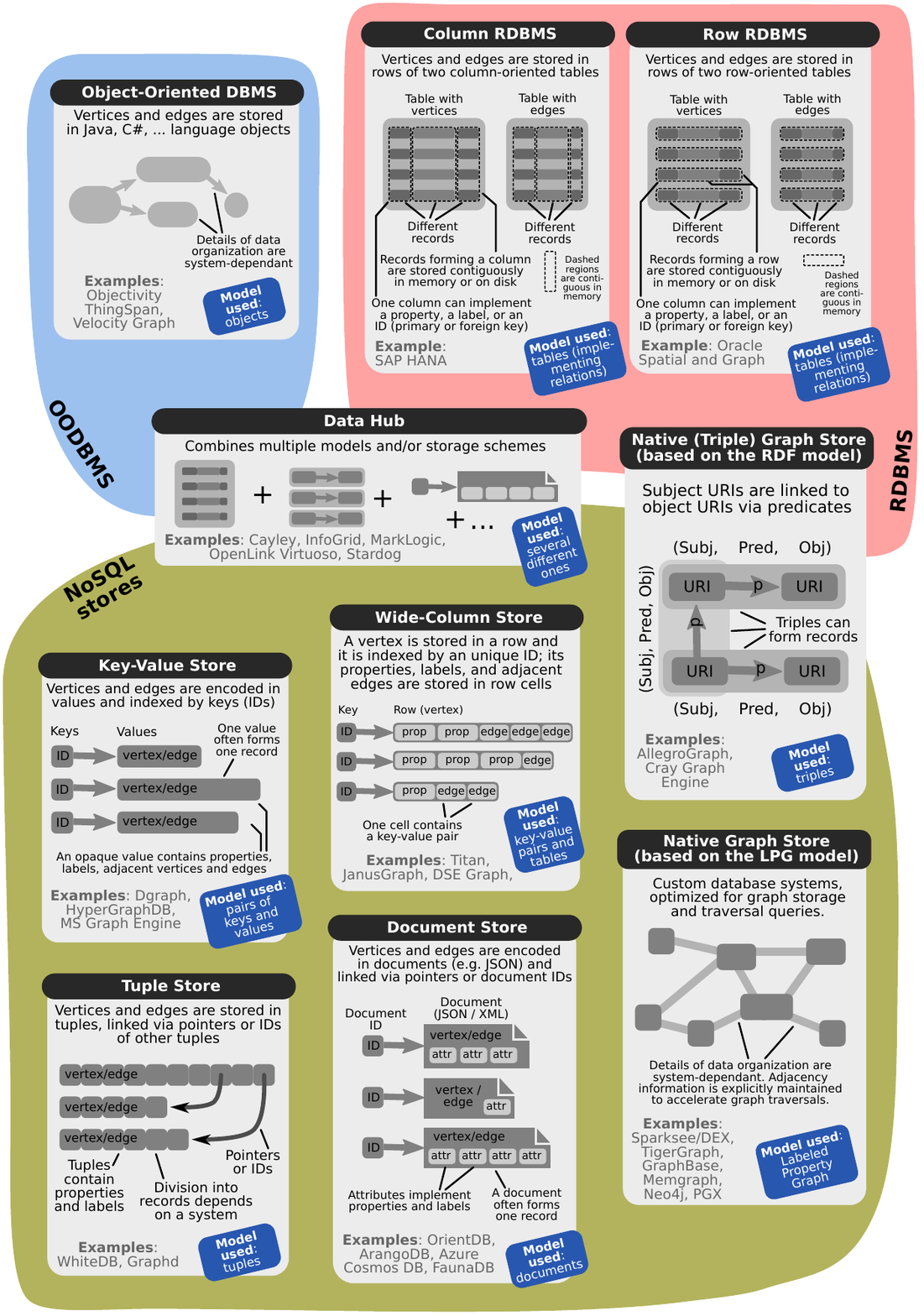}
\vspace{-2.5em}
\caption{
Overview of different \textbf{graph storage backends}, with examples.
}
\label{fig:categories}
\end{figure*}

\iftr
We now describe how we categorize \emph{graph database systems} considered in this
survey~\cite{allegro_graph_links,
amazon_neptune_links, anzo_graph_links, mormotta_links, blaze_graph_links,
brightstardb_links, graphdb_links, profium_sense_links, triplebit_links,
graphd_links, azure_cosmosdb_links, bitsy_links, faunadb_links, dgraph_links,
redisgraph_links, datastax_links, hgraphdb_links, titan_link,
graphgen_rdbms_paper, janus_graph_links, velocitygraph_links, neo4j_book,
sparksee_paper, gbase_paper, virtuoso_links, marklogic_links, cge_paper,
whitedb_links, orientdb_links, arangodb_links, arangodb_indexing_links,
trinity_paper, hypergraphdb_links}. 
\fi
This taxonomy incorporates existing concepts related to graph data models
(cf.~\cref{graph_models_section}) and to graph queries
(cf.~\cref{sec:queries-workloads}). Then, other aspects of the
proposed taxonomy are novel. In this section, we describe the taxonomy in a
general way. In \cref{sec:main-discussion}, we analyze the taxonomy in
the context of specific graph database systems.
The main dimensions of the taxonomy are (1) the general backend type, (2) data
organization, and (3) query execution.
Figure~\ref{fig:categories} illustrates the general types of considered
databases together with certain aspects of data models and organization.
Figure~\ref{fig:taxonomy} summarizes all elements of the proposed
taxonomy.

\iftr
\begin{figure*}[htbp]
\else
\begin{figure*}[t]
\fi
\centering
\includegraphics[width=1.0\textwidth]{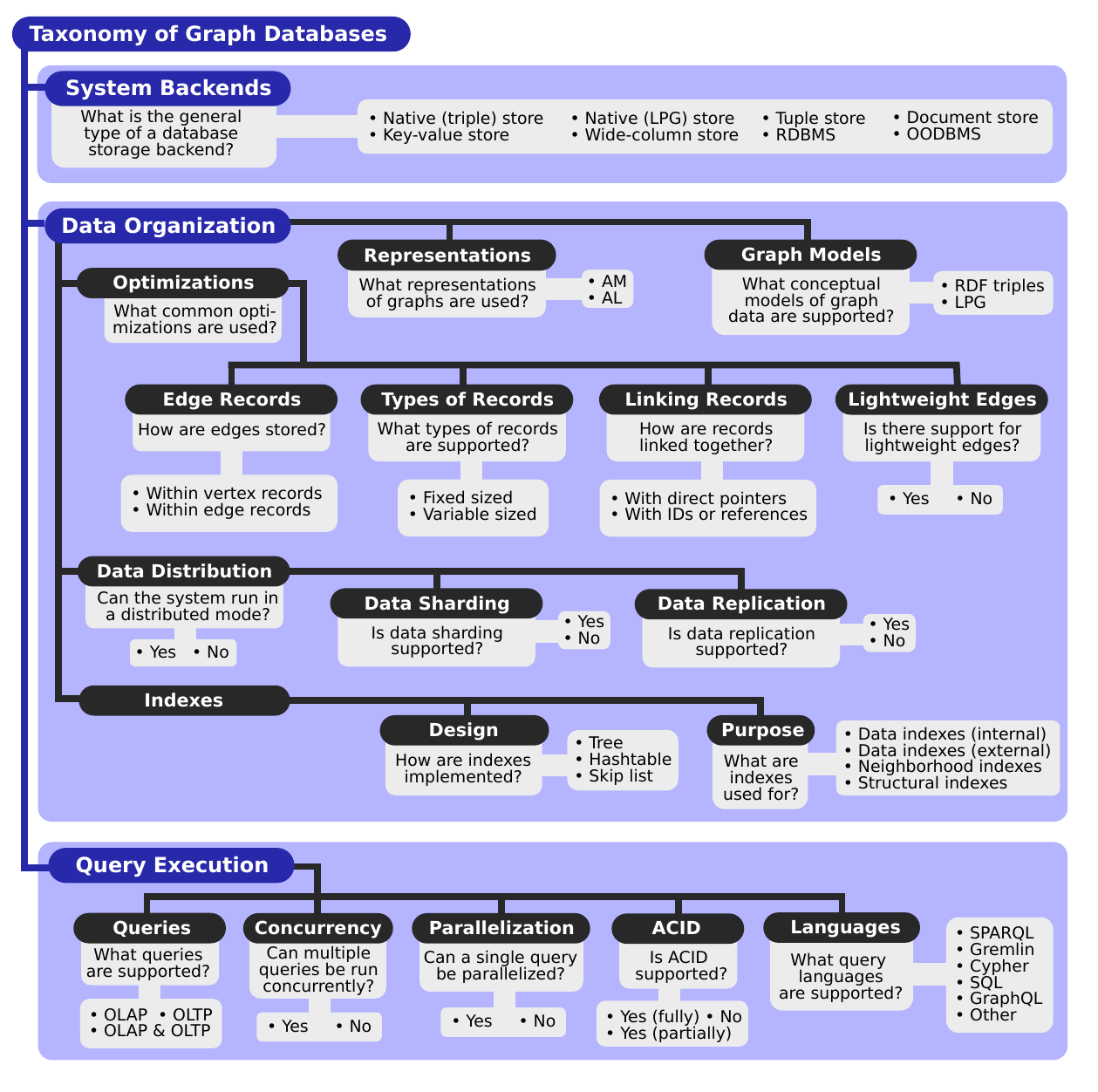}
\vspace{-2em}
\caption{Overview of the \textbf{identified taxonomy of graph databases}.
}
\label{fig:taxonomy}
\end{figure*}

\subsection{Types of Graph Database Storage Backends}
\label{sec:taxonomy_types}

We first identify \textbf{general types of graph databases} that primarily
differ in their \textbf{storage backends}
(e.g., a triple store or a document store). This facilitates further
taxonomization and analysis because (1) the backend design has a profound
impact on almost all other aspects of a graph database such as data
organization, and because (2) it straightforwardly enables categorizing all
considered graph databases into a few clearly defined groups.

Some classes of systems use a certain \emph{specific backend}
technology, adapting this backend to storing graph data, and adding a
\emph{frontend} to query the graph data. Examples of such systems are
\textbf{tuple} stores, \textbf{document} stores, \textbf{key-value} stores,
\textbf{wide-column} stores, \textbf{Relational} Database Management Systems
(RDBMS), or \textbf{Object-Oriented} Database Management Systems (OODBMS). 
Other graph databases are designed \emph{specifically} for maintaining and
querying graphs; we call such systems \textbf{native graph databases} (or
\textbf{native graph stores}), they are based on either the \textbf{LPG} or
the \textbf{RDF} graph data model.
Finally, we consider designs called the \textbf{data hubs}; they enable using
\emph{many different} storage backends, facilitating storing data in different
formats and models.

Some of the above categories of systems fall into the domain of NoSQL
stores. For example, this includes document stores, key-value stores, or some
triple stores. However, there is no strict assignment of specific storage
backends as NoSQL. For example, triple stores can also be implemented as, e.g.,
RDBMS~\cite{davoudian2018survey}.
Figure~\ref{fig:categories} illustrates these systems, they are
discussed in more detail in~\cref{existing_gdb}.

\subsection{Data Organization}

In data organization, we distinguish the following taxonomy aspects: (1) graph
structure representation, (2) conceptual data models, (3) indexes, (4) data
distribution, and (5) common optimizations.

First, in \textbf{graph structure representation}, we analyze whether the graph
structure is stored using the AL or the AM representation
(see~\cref{sec:g_reps}). The graph structure representation directly
impacts the performance of queries.

Second, we investigate what \textbf{conceptual data models} are supported by
different graph databases. Here, we focus on the RDF and LPG models as well as
their variants, described in~\cref{sec:g_models}. The used graph model
strongly influences what graph query languages can be used together with a
given system, and it also has impact on the associated data layout.

We also analyze \textbf{how graph databases use indexes} to accelerate
accessing data. Indexes can significantly improve the
performance of GDBs, and they are widely used, for example in RDF
systems~\cite{abburu2015effective, punnoose2012rya, khadilkar2012jena}.
Here, we consider the functionality (i.e., the use case) of a
given index, and how a given index is implemented \footnote{We do not include
the index information in Tables~\ref{survey-table1}--\ref{survey-table2}
because of lack of space, and instead provide a detailed separate analysis
in~\cref{sec:dis-indexes}.}.
As for the former, we identify four different index use cases: storing the
locations of vertex neighborhoods (referred to as ``neighborhood indexes''),
indexing graph elements, such as vertices, that satisfy pre-specified conditions
related to rich graph data (referred to as ``data indexes''),
storing the actual graph data, and maintaining non-graph related data (referred
to as ``structural indexes'').
As for the latter, we identify three fundamental data structures used to
implemented indexes: trees, skip lists, and hashtables.

We also identify whether a database can run in a \textbf{distributed mode}.
A system is \textit{distributed} or \textit{multi-server} if it can run on
multiple servers (also called compute nodes) connected with a network. 
In such systems, data may be \textit{replicated}~\cite{replication_book}
(maintaining copies of the dataset at each server), or it may allow for
\textit{sharding}~\cite{fragmentation_book} (data fragmentation, i.e., storing
only a part of the given dataset on one server). Replication often allows for
more fault tolerance~\cite{advantage_distrDB_book}, sharding reduces the amount
of used memory per node and can improve
performance~\cite{advantage_distrDB_book}.
Gathering this information facilitates selecting a system with the most
appropriate performance properties in a given context. For example, systems
that replicate but not shard the data, may offer more performance for read-only
workloads, but may scale badly for particularly large graphs that would
require disk spilling.

Finally, we identify \textbf{common optimizations}:
(1) dividing data into records,
(2) lightweight edges, and
(3) linking records with direct pointers.
%

\textbf{Dividing Data into Records}
Graph databases usually organize data into small units called
\textit{records}. One record contains information about a certain single
entity (e.g., a person), this information is organized into specified logical
fields (a name, a surname, etc.). A certain number of records is often
kept together in one contiguous block in memory or disk to enhance data access
locality.

\textbf{Enabling Lightweight Edges}
Some systems (e.g., OrientDB) allow edges without labels or properties to be
stored as \textit{lightweight edges}.  Such edges are stored in the records of
the corresponding source and/or destination vertices.  These
lightweight edges are represented by the ID of their destination
vertex, or by a pointer to this vertex.  This can save storage space and
accelerate resolving different graph queries such as verifying connectivity of
two vertices~\cite{orientdb_lwedge_links}.

\textbf{Linking Records with Direct Pointers}
In record based systems, vertices and edges are stored in records.  To enable
efficient resolution of connectivity queries (i.e., verifying whether two
vertices are connected), these records have to point to other records.  One
option is to store \textit{direct pointers} (i.e., memory addresses) to the
respective connected records. For example, an edge record can store direct
pointers to vertex records with adjacent vertices.  Another option is to assign
each record a unique ID and use these IDs instead of direct pointers to refer
to other records.  On one hand, this requires an additional indexing structure
to find the physical location of a record based on its ID.  On the other hand,
if the physical location changes, it is usually easier to update the indexing
structure instead of changing all associated direct pointers.

Note that specific systems can employ other diverse optimizations. For example,
in addition to using index structures to maintain the locations of data, some
databases also store the graph data in the indexes themselves. In such cases,
the index does not point to a certain data record but the index itself contains
the desired data. Example systems with such functionality are Sparksee/DEX and
Cray Graph Engine. To maintain indices, the former uses bitmaps and B+ trees
while the latter uses hashtables.

\subsection{Query Execution}
\label{sec:taxonomy_execution}

In query execution, we identify the following aspects: (1) concurrent execution
of different queries, (2) parallelization of single queries, (3) ACID
transactions, (4) support for classes of queries, and (5) support for query
languages.

Note that almost all of the studied graph databases are closed source or do
not come with any associated discussions of the details of the design of query
execution (except for general descriptions). Thus, we do not offer a detailed
associated taxonomy for algorithmic aspects of query execution, beyond the
above criteria. However, we provide a detailed associated discussion on a few
systems that do come with more details on their query execution. We also
analyze relationships between the backend type, the data organization, and the
query execution. This enables deriving certain insights about the design of
different backends. For example, the query language support is primarily
affected by the supported conceptual graph model; if it is RDF, then the system
usually supports SPARQL while systems focusing on LPG usually support Cypher or
Gremlin.


We define \textbf{concurrent execution} as the execution of separate queries at
the same time. Concurrent execution of queries can lead to higher throughput.
%
%
We also define \textbf{parallel execution} as the parallelized execution of a
single query, possibly on more than one server or compute node. Parallel
execution can lead to lower latencies for queries that can be parallelized.
\iftr
In~\cref{sec:discussion-query-exec}, we correlate
the support for concurrent and parallel queries with different fundamental backend types,
and we describe the details of query execution in graph databases that disclose this
information.
\fi


Many graph databases support \textit{transactions}; we analyze them in~\cref{sec:discussion-transact}.
%
%
\textbf{ACID}~\cite{hoffer2011modern} (Atomicity, Consistency, Isolation,
Durability) is a well-known set of properties that database transactions uphold
in many database systems. Different graph databases explicitly ensure 
some or all of ACID.


We also analyze \textbf{supported queries}
in~\cref{sec:discussion-queries}.
Some databases (e.g., ArangoDB~\cite{arangodb_links}) are oriented
towards OLTP, where focus is on executing many smaller, interactive,
transactional queries.
%
%
Other systems (e.g., Cray Graph Engine~\cite{cge_paper}) focus more on
OLAP: they execute analytics queries that span the whole graphs, usually taking
more time than OLTP.
Finally, different databases (e.g., Neo4j~\cite{neo4j_book}) offer
extensive support for both.
%

Although we do not focus on graph database languages, we also report on
\textbf{supported query languages} in~\cref{sec:discussion-languages}). We
consider the leading languages such as SPARQL~\cite{perez2009semantics},
Gremlin~\cite{rodriguez2015gremlin}, Cypher~\cite{francis2018cypher,
holzschuher2013performance, green2019updating}, and SQL~\cite{date1987guide}.
We also mention other system-specific languages such as
GraphQL~\cite{hartig2018semantics} and support for APIs from languages such as
C++ or Java\footnote{We bring to the reader's attention a manifesto on creating
GQL, a standardized graph query language (\url{https://gql.today}).}.
Note that mapping graph queries to SQL was also addressed in past
work~\cite{steer2017cytosm}.

\begin{table}[hbtp]
%
\setlength{\tabcolsep}{0.8pt}
\renewcommand{\arraystretch}{0.9}
\centering
\ssmall
\sf
\begin{tabular}{lllllllllllllllllllll}
\toprule
\multirow{2}{*}{\makecell[l]{\textbf{Graph Database}\\ \textbf{System}}}
& \multirow{2}{*}{\makecell[l]{\textbf{oB}}}
& \multicolumn{2}{c}{Model} 
& \multicolumn{2}{c}{Repr.}
& \multicolumn{6}{c}{Data Organization} 
& \multicolumn{8}{c}{Data Distribution \& Query Execution} & \multirow{2}{*}{\textbf{Additional
remarks}} \\
\cmidrule(lr){3-4}
\cmidrule(lr){5-6}
\cmidrule(lr){7-12}
\cmidrule(lr){13-20}
%
 & & \textbf{lpg} & \textbf{rdf} & \textbf{al} & \textbf{am} & \textbf{fs} & \textbf{vs} & \textbf{dp} & \textbf{se} & \textbf{sv}& \textbf{lw} & \textbf{ms} & \textbf{rp} & \textbf{sh} & \textbf{ce} & \textbf{pe} & \textbf{tr} & \textbf{oltp} & \textbf{olap} & \\
\midrule
\multicolumn{21}{l}{\textbf{NATIVE GRAPH DATABASES (RDF model based, triple stores)} (\cref{triple_store_section}). The main data model used: \textbf{RDF triples} (\cref{rdf_section}).} \\
\midrule 
\textbf{AllegroGraph \cite{allegro_graph_links}}     & \faThumbsODown & \faThumbsODown & \faThumbsUp  & \faThumbsODown & \faThumbsODown & \faThumbsUp$^{*}$ & \faThumbsODown & \faThumbsODown  & \faThumbsODown & \faThumbsODown & \faThumbsODown & \faThumbsUp & \faThumbsUp & \faThumbsUp & \faThumbsUp & \faThumbsODown & \faThumbsUp & \faThumbsUp & \noAnswer & $^{*}$Triples are stored as integers (RDF strings\\
\commentline{map to integers).}\\ 
\textbf{BlazeGraph~\cite{blaze_graph_links}} & \faThumbsODown & \faThumbsUp$^{*}$ & \faThumbsUp$^{*}$ & \faThumbsODown &  \faThumbsODown & \noAnswer & \noAnswer & \faThumbsODown  & \faThumbsODown & \faThumbsODown & \faThumbsODown & \faThumbsUp & \faThumbsUp & \faThumbsUp & \noAnswer & \noAnswer & \faThumbsUp & \noAnswer & \noAnswer & $^{*}$BlazeGraph uses RDF*, an extension of RDF\\
\commentline{(details in~\cref{ab_section}).}\\
\textbf{Cray Graph Engine~\cite{cge_paper}}      & \faThumbsODown &\faThumbsODown & \faThumbsUp  & \faThumbsODown &  \faThumbsODown & \faThumbsODown$^{*}$& \faThumbsODown$^{*}$& \faThumbsODown    & \faThumbsODown & \faThumbsODown & \faThumbsODown & \faThumbsUp & \faThumbsODown  & \faThumbsUp & \faThumbsODown & \faThumbsUp & \faThumbsODown & \faThumbsODown & \faThumbsUp &$^{*}$RDF triples are stored in hashtables. \\ 
Amazon Neptune \cite{amazon_neptune_links}  & \faThumbsODown & \faThumbsUp & \faThumbsUp  & \faThumbsODown  & \faThumbsODown & \noAnswer & \noAnswer & \faThumbsODown & \faThumbsODown & \faThumbsODown& \faThumbsODown & \faThumbsUp & \faThumbsUp & \faThumbsODown & \faThumbsUp & \faThumbsODown & \faThumbsUp & \faThumbsUp & \faThumbsUp & --- \\
AnzoGraph \cite{anzo_graph_links}           & \faThumbsODown & \faThumbsUp & \faThumbsUp   & \faThumbsODown & \faThumbsODown & \noAnswer & \noAnswer & \faThumbsODown & \faThumbsODown & \faThumbsODown& \faThumbsODown & \faThumbsUp & \faThumbsODown & \faThumbsUp & \faThumbsUp & \faThumbsUp & \faThumbsUp & \faThumbsUp & \faThumbsUp & --- \\
Apache Jena TBD~\cite{apache_jena_tbd_links} & \faThumbsODown & \faThumbsODown  & \faThumbsUp  & \noAnswer &\faThumbsODown & \noAnswer & \noAnswer & \noAnswer  & \noAnswer & \noAnswer & \noAnswer & \faThumbsODown & \noAnswer & \faThumbsODown & \faThumbsUp & \faThumbsODown & \faThumbsUp & \faThumbsUp & \noAnswer & --- \\
Apache Marmotta~\cite{mormotta_links} & \faThumbsODown & \faThumbsODown  & \faThumbsUp  &  \faThumbsODown  &\faThumbsODown & \faThumbsUp$^{*}$ & \faThumbsODown & \faThumbsODown & \faThumbsODown & \faThumbsODown & \faThumbsODown & \noAnswer & \noAnswer & \noAnswer & \faThumbsUp & \faThumbsUp & \faThumbsUp & \faThumbsUp & \faThumbsUp & $^{*}$The structure of data records is based on\\ \commentline{that of different RDBMS systems}\\ \commentline{(H2~\cite{mueller2006h2}, PostgreSQL~\cite{momjian2001postgresql}, MySQL~\cite{dubois1999mysql}).} \\
BrightstarDB~\cite{brightstardb_links} & \faThumbsODown & \faThumbsODown  & \faThumbsUp  & \faThumbsODown & \faThumbsODown & \noAnswer & \noAnswer & \faThumbsODown & \faThumbsODown & \faThumbsODown & \faThumbsODown & \noAnswer & \noAnswer & \noAnswer & \faThumbsUp & \noAnswer & \faThumbsUp & \faThumbsUp & \noAnswer & --- \\
gStore~\cite{gstore} & \faThumbsODown & \faThumbsODown  & \faThumbsUp  & \faThumbsUp & \faThumbsODown& \faThumbsODown & \faThumbsUp & \faThumbsODown & \faThumbsODown & \faThumbsUp & \faThumbsODown & \noAnswer & \noAnswer & \faThumbsODown & \noAnswer & \faThumbsODown & \noAnswer & \noAnswer & \noAnswer & --- \\
Ontotext GraphDB~\cite{graphdb_links} & \faThumbsODown & \faThumbsODown  & \faThumbsUp & \faThumbsODown  & \faThumbsODown & \noAnswer & \noAnswer & \faThumbsODown & \faThumbsODown & \faThumbsODown & \faThumbsODown & \faThumbsUp & \faThumbsUp & \faThumbsODown & \faThumbsUp & \noAnswer & \faThumbsUp & \faThumbsUp & \noAnswer & --- \\
Profium Sense~\cite{profium_sense_links} & \faThumbsODown & \faThumbsODown   & \faThumbsUp$^*$  & \faThumbsODown & \faThumbsODown & \noAnswer & \noAnswer & \faThumbsODown & \faThumbsODown & \faThumbsODown& \faThumbsODown & \faThumbsUp & \faThumbsUp & \noAnswer & \faThumbsUp & \noAnswer & \faThumbsUp & \faThumbsUp & \noAnswer & $^{*}$The format used is called JSON-LD:\\ \commentline{JSON for vertices and RDF for edges.} \\
TripleBit~\cite{triplebit_links} & \faThumbsODown & \faThumbsODown   & \faThumbsUp  & \faThumbsODown  & \faThumbsODown & \faThumbsODown & \faThumbsUp$^{*}$ & \faThumbsODown & \faThumbsODown & \faThumbsODown& \faThumbsODown & \faThumbsODown$^{\ddagger}$ & \faThumbsODown & \faThumbsODown & \faThumbsODown & \faThumbsODown & \noAnswer & \noAnswer & \faThumbsUp & The data organization uses compression.\\ \commentline{$^{*}$Strings map to variable size integers.}\\ \commentline{$^{\ddagger}$Described as future work.}\\
\midrule
\multicolumn{21}{l}{\textbf{NATIVE GRAPH DATABASES (LPG model based)} (\cref{nativegdb_section}). The main data model used: \textbf{LPG} (\cref{lpg_defininion}, \cref{lpgvariants_section}).} \\
\midrule
%
%
\textbf{Neo4j~\cite{neo4j_book}}      & \faThumbsODown & \faThumbsUp & \faThumbsODown   & \faThumbsUp  &  \faThumbsODown & \faThumbsUp   & \faThumbsODown & \faThumbsUp         & \faThumbsUp   & \faThumbsODown & \faThumbsODown & \faThumbsUp & \faThumbsUp & \faThumbsODown & \faThumbsUp & \faThumbsODown & \faThumbsUp & \faThumbsUp & \faThumbsUp & Neo4j is provided as a cloud service by a\\
\commentline{system called Graph Story~\cite{graph_story_links}}. \\
\textbf{Sparksee/DEX~\cite{sparksee_paper}} & \faThumbsODown & \faThumbsUp  & \faThumbsODown   &\faThumbsODown$^{*}$  & \faThumbsODown &\faThumbsODown$^{\ddagger}$&\faThumbsODown$^{\ddagger}$& \faThumbsODown & \faThumbsODown & \faThumbsODown & \faThumbsODown & \faThumbsUp & \faThumbsUp & \faThumbsODown & \faThumbsUp & \faThumbsUp & \faThumbsUp & \faThumbsUp & \faThumbsUp &$^{*}$Bitmaps are used for connectivity.\\\commentline{$^{\ddagger}$The system uses maps only.}\\
GBase~\cite{gbase_paper} & \faThumbsODown &\faThumbsODown$^{*}$ & \faThumbsODown  &\faThumbsODown$^{\ddagger}$ & \faThumbsUp  & \faThumbsODown & \faThumbsODown & \faThumbsODown & \faThumbsODown & \faThumbsODown & \faThumbsODown & \faThumbsUp & \noAnswer & \noAnswer & \noAnswer & \noAnswer & \noAnswer & \faThumbsODown & \faThumbsUp &$^{*}$GBase supports simple graphs only (\cref{simplegraph_section}).\\
\commentline{$^{\ddagger}$GBase stores the AM sparsely.} \\
GraphBase~\cite{graphbase_links} & \faThumbsODown & \faThumbsODown$^{*}$  & \faThumbsODown  & \noAnswer & \faThumbsODown & \faThumbsODown & \faThumbsUp & \noAnswer & \noAnswer & \noAnswer& \noAnswer & \faThumbsUp & \noAnswer & \faThumbsUp & \faThumbsUp & \noAnswer & \faThumbsUp & \faThumbsUp & \noAnswer & $^{*}$No support for edge properties, only two\\
\commentline{types of edges available.} \\
Graphflow~\cite{graphflow} & \faThumbsODown & \faThumbsUp  & \faThumbsODown  & \faThumbsUp  & \faThumbsODown & \noAnswer & \noAnswer & \noAnswer & \noAnswer & \noAnswer & \noAnswer & \faThumbsODown & \noAnswer & \noAnswer & \noAnswer & \noAnswer & \noAnswer & \noAnswer & \faThumbsUp & --- \\
LiveGraph~\cite{livegraph} & \faThumbsODown & \faThumbsUp  & \faThumbsODown  & \faThumbsUp & \faThumbsODown & \faThumbsODown & \faThumbsUp & \faThumbsODown & \faThumbsUp & \faThumbsODown & \faThumbsODown & \faThumbsODown & \noAnswer & \faThumbsODown & \faThumbsUp & \noAnswer & \faThumbsUp & \faThumbsUp & \faThumbsUp & --- \\
Memgraph~\cite{memgraph_links} & \faThumbsODown & \faThumbsUp & \faThumbsODown  & \faThumbsUp & \faThumbsODown & \noAnswer & \noAnswer & \noAnswer & \noAnswer & \noAnswer& \noAnswer & \faThumbsUp & \faThumbsUp & \faThumbsUp & \faThumbsOUp$^{*}$ & \faThumbsOUp$^{\ddagger}$ & \faThumbsUp & \faThumbsUp & \faThumbsUp & $^{*}$This feature is under development.\\ \commentline{$^{\ddagger}$Available only for some algorithms.}\\
TigerGraph~\cite{tiger_graph_links} & \faThumbsODown & \faThumbsUp  & \faThumbsODown  & \noAnswer & \faThumbsODown & \noAnswer & \noAnswer & \noAnswer  & \noAnswer & \noAnswer& \noAnswer & \faThumbsUp & \faThumbsUp & \faThumbsUp & \faThumbsUp & \faThumbsUp & \faThumbsUp & \faThumbsUp & \faThumbsUp & --- \\
Weaver~\cite{dubey2016weaver} & \faThumbsODown & \faThumbsUp  & \faThumbsODown & \noAnswer & \faThumbsODown & \noAnswer & \noAnswer & \noAnswer & \noAnswer & \noAnswer & \noAnswer & \faThumbsUp & \faThumbsUp & \faThumbsUp & \faThumbsUp & \faThumbsUp & \faThumbsUp & \faThumbsUp & \faThumbsUp & --- \\
\midrule
\multicolumn{21}{l}{\textbf{KEY-VALUE STORES} (\cref{keyvalue_section}). The main data model used: \textbf{key-value pairs} (\cref{kv_section}).} \\
\midrule
\textbf{HyperGraphDB~\cite{hypergraphdb_links}} & \faThumbsODown &\faThumbsODown$^{*}$  & \faThumbsODown  &\faThumbsODown$^{\ddagger}$  & \faThumbsODown & \faThumbsODown & \faThumbsUp   & \faThumbsODown & \faThumbsUp   & \faThumbsODown & \faThumbsODown & \faThumbsUp & \faThumbsUp & \faThumbsUp & \faThumbsUp & \faThumbsUp & \faThumbsOUp$^{\dagger}$ & \faThumbsUp & \faThumbsUp &$^{*}$A Hypergraph model. $^{\ddagger}$The system uses\\
\commentline{an incidence index to retrieve edges of a}\\
\commentline{vertex. $^{\dagger}$Support for ACI only.} \\
\textbf{MS Graph Engine~\cite{trinity_paper}} & \faThumbsUp   & \faThumbsUp   & \faThumbsODown  &\faThumbsUp$^{*}$  & \faThumbsODown  & \faThumbsODown &\faThumbsUp$^{\ddagger}$& \faThumbsODown & \faThumbsUp   & \faThumbsODown & \faThumbsUp   & \faThumbsUp & \faThumbsODown & \faThumbsUp & \faThumbsUp & \faThumbsUp & \faThumbsODown & \faThumbsUp & \faThumbsUp & $^{*}$AL contains IDs of edges and/or vertices.\\ \commentline{$^{\ddagger}$Schema is defined by Trinity} \\ \commentline{Specification Language (TSL).} \\
Dgraph \cite{dgraph_links} & \faThumbsODown & \faThumbsUp  & \faThumbsODown  & \faThumbsUp & \faThumbsODown & \faThumbsODown & \faThumbsUp & \faThumbsODown & \faThumbsUp & \faThumbsODown& \faThumbsODown & \faThumbsUp & \faThumbsUp & \faThumbsUp & \faThumbsUp & \faThumbsUp & \faThumbsUp & \faThumbsUp & \faThumbsUp & Dgraph is based on Badger~\cite{badger_links}.\\
RedisGraph~\cite{redisgraph_links, rana_redisgraph, dbengines_redisgraph} & \faThumbsODown & \faThumbsUp  & \faThumbsODown  & \faThumbsODown & \faThumbsODown & \faThumbsODown & \faThumbsUp & \faThumbsODown & \faThumbsODown & \faThumbsODown & \faThumbsODown & \faThumbsUp & \faThumbsUp & \faThumbsODown & \faThumbsUp & \faThumbsODown & \faThumbsODown & \faThumbsODown & \faThumbsUp$^{*}$ & RedisGraph is based on Redis~\cite{redis_links}.\\ \commentline{$^{*}$The OLAP part uses GraphBLAS~\cite{kepner2016mathematical}.}\\
%
%
\midrule
\multicolumn{21}{l}{\textbf{DOCUMENT STORES} (\cref{document_store_section}). The main data model used: \textbf{documents} (\cref{doc_section}).} \\
\midrule
\textbf{ArangoDB~\cite{arangodb_links}} & \faThumbsUp   & \faThumbsUp  & \faThumbsODown  &\faThumbsODown$^{*}$ & \faThumbsODown & \faThumbsODown & \faThumbsUp   & \faThumbsODown & \faThumbsUp   & \faThumbsODown & \faThumbsODown & \faThumbsUp & \faThumbsUp & \faThumbsUp & \faThumbsUp & \faThumbsODown & \faThumbsUp & \faThumbsUp & \faThumbsOUp & $^{*}$Uses a hybrid index for retrieving edges. \\
\textbf{OrientDB~\cite{orientdb_links}} & \faThumbsUp   & \faThumbsUp  & \faThumbsODown  &\faThumbsUp$^{*}$  & \faThumbsODown  & \faThumbsODown & \faThumbsUp   & \faThumbsUp   & \faThumbsUp   & \faThumbsODown & \faThumbsUp   & \faThumbsUp & \faThumbsUp & \faThumbsUp$^{\ddagger}$ & \faThumbsUp & \faThumbsUp & \faThumbsUp & \faThumbsUp & \faThumbsODown & $^{*}$AL contains RIDs (i.e., physical locations)\\ \commentline{of edge and vertex records. $^{\ddagger}$Sharding is}\\ \commentline{user defined. OrientDB supports JSON and}\\ \commentline{it offers certain object-oriented capabilities.}\\
Azure Cosmos DB \cite{azure_cosmosdb_links} & \faThumbsUp & \faThumbsUp & \faThumbsODown   & \faThumbsODown & \faThumbsODown & \faThumbsODown & \faThumbsUp & \faThumbsODown & \faThumbsUp & \faThumbsODown & \faThumbsODown & \faThumbsUp & \faThumbsUp & \faThumbsUp & \faThumbsUp & \faThumbsODown & \faThumbsUp & \faThumbsUp & \noAnswer & ---  \\
Bitsy~\cite{bitsy_links} & \faThumbsODown & \faThumbsUp   & \faThumbsODown  & \faThumbsODown   & \faThumbsODown & \faThumbsODown & \faThumbsUp & \faThumbsODown & \faThumbsUp & \faThumbsODown& \faThumbsODown & \faThumbsODown & \faThumbsODown & \faThumbsODown & \faThumbsUp & \faThumbsODown & \faThumbsUp & \faThumbsUp & \faThumbsODown & The system is disk based and uses JSON files.\\ \commentline{The storage only allows for appending data}. \\
FaunaDB~\cite{faunadb_links} & \faThumbsUp$^{*}$ & \faThumbsUp  & \faThumbsODown  & \faThumbsUp$^{\ddagger}$ & \faThumbsODown & \faThumbsODown & \faThumbsUp & \faThumbsUp  & \faThumbsUp & \faThumbsODown & \faThumbsODown & \faThumbsUp & \faThumbsUp & \faThumbsUp & \faThumbsUp & \faThumbsODown & \faThumbsUp & \faThumbsUp & \faThumbsODown & $^{*}$Document, RDBMS, graph, ``time series''. \\ \commentline{$^{\ddagger}$Adjacency lists are separately precomputed.}\\
%
%
\bottomrule
\end{tabular}
%
\caption{
\textbf{Comparison of graph databases {(native graph databases based on RDF and LPG, key-value stores, and document stores)}}.
\textbf{\ul{Bolded systems}} are described in more detail in the corresponding
sections.
\textbf{\ul{oB}}: A system supports secondary data models / backend types (in addition to its primary one).
\textbf{\ul{lpg}}, \textbf{\ul{rdf}}: A system supports, respectively, the
\textbf{Labeled Property Graph} and \textbf{RDF} without prior data
transformation.
\textbf{\ul{am}, \ul{al}}: The structure is represented as the \textbf{adjacency matrix} or the \textbf{adjacency list}.
\textbf{\ul{fs}}, \textbf{\ul{vs}}: Data records are \textbf{fixed size} and
\textbf{variable size}, respectively.
\textbf{\ul{dp}}: A system can use \textbf{direct
pointers} to link records. This enables {storing and traversing adjacency data without
maintaining indices}.
\textbf{\ul{se}}: Edges can be \textbf{stored in a separate edge record}.
\textbf{\ul{sv}}: Edges can be \textbf{stored in a vertex record}.
\textbf{\ul{lw}}: Edges can be \textbf{lightweight} (containing just a vertex ID or a pointer, both stored in a vertex record).
\textbf{\ul{ms}}: A system can operate in a \textbf{Multi-Server} (distributed) mode.
\textbf{\ul{rp}}: Given a distributed mode, a system enables \textbf{Replication} of datasets.
\textbf{\ul{sh}}: Given a distributed mode, a system enables \textbf{Sharding} of datasets.
\textbf{\ul{ce}}: Given a distributed mode, a system enables \textbf{Concurrent Execution} of multiple queries.
\textbf{\ul{pe}}: Given a distributed mode, a system enables \textbf{Parallel Execution} of single queries on multiple nodes/CPUs.
\textbf{\ul{tr}}: Support for \textbf{ACID Transactions}.
\textbf{\ul{oltp}}: Support for \textbf{Online Transaction Processing}.
\textbf{\ul{olap}}: Support for \textbf{Online Analytical Processing}.
\faThumbsUp: A system offers a given feature.
\faThumbsOUp: A system offers a given feature in a limited way.
\faThumbsODown: A system does not offer a given feature.
\noAnswer: Unknown.\\
}
\label{survey-table1}
%
%
\end{table}

\begin{table}[hbtp]
\setlength{\tabcolsep}{0.8pt}
\renewcommand{\arraystretch}{0.8}
\centering
\ssmall
\sf
\begin{tabular}{lllllllllllllllllllll}
\toprule
\multirow{2}{*}{\makecell[l]{\textbf{Graph Database}\\ \textbf{System}}}
& \multirow{2}{*}{\makecell[l]{\textbf{oB}}}
& \multicolumn{2}{c}{Model}
& \multicolumn{2}{c}{Repr.}
& \multicolumn{6}{c}{Data Organization}
& \multicolumn{8}{c}{Data Distribution \& Query Execution}
& \multirow{2}{*}{\textbf{Additional remarks}} \\
\cmidrule(lr){3-4}
\cmidrule(lr){5-6}
\cmidrule(lr){7-12}
\cmidrule(lr){13-20}
%
 & & \textbf{lpg} & \textbf{rdf} & \textbf{al} & \textbf{am} & \textbf{fs} & \textbf{vs} & \textbf{dp} & \textbf{se} & \textbf{sv}& \textbf{lw} & \textbf{ms} & \textbf{rp} & \textbf{sh} & \textbf{ce} & \textbf{pe} & \textbf{tr} & \textbf{oltp} & \textbf{olap} & \\
\midrule
\multicolumn{21}{l}{\textbf{RELATIONAL DBMS (RDBMS)} (\cref{rdbms_section}). The main data model used: \textbf{tables (implementing relations)} (\cref{table_section}).} \\
\midrule
\textbf{Oracle Spatial} & \faThumbsUp$^{*}$ & \faThumbsUp & \faThumbsUp & \faThumbsODown & \faThumbsODown & \noAnswer$^{*}$ & \noAnswer$^{*}$ & \faThumbsODown & \faThumbsUp$^{*}$ & \faThumbsODown & \faThumbsODown & \faThumbsUp & \faThumbsUp & \faThumbsUp & \faThumbsUp & \faThumbsUp & \faThumbsUp & \faThumbsUp & \faThumbsUp & $^{*}$LPG and RDF use row-oriented storage.\\
\textbf{and Graph~\cite{oracle_spatial}} & & & & & & & & & & & & & & & & & & & & The system can also run on top of PGX~\cite{hong2015pgx}\\
\commentline{(effectively as a \textbf{native graph database}).} \\
AgensGraph~\cite{agensgraph_links} & \faThumbsUp & \faThumbsUp & \faThumbsODown & \faThumbsODown & \faThumbsODown & \noAnswer & \noAnswer & \faThumbsODown & \noAnswer & \faThumbsODown & \faThumbsODown & \faThumbsUp & \faThumbsUp & \faThumbsUp & \faThumbsUp & \faThumbsUp & \faThumbsUp & \faThumbsUp & \faThumbsUp & AgensGraph is based on PostgreSQL. \\
FlockDB~\cite{flockdb_links} & \faThumbsODown & \faThumbsODown & \faThumbsODown & \faThumbsODown & \faThumbsODown & \noAnswer & \noAnswer & \faThumbsODown & \noAnswer & \faThumbsODown& \faThumbsODown & \faThumbsUp & \faThumbsUp & \faThumbsUp & \faThumbsUp & \faThumbsODown & \faThumbsODown & \faThumbsUp & \faThumbsODown & The system focuses on ``shallow'' graph\\
\commentline{queries, such as finding mutual friends.} \\
IBM Db2 & \faThumbsODown & \faThumbsUp & \faThumbsODown & \faThumbsODown & \faThumbsODown & \noAnswer & \noAnswer & \faThumbsODown & \faThumbsUp & \faThumbsUp$^{*}$ & \faThumbsUp$^{*}$ & \faThumbsUp$^{\ddagger}$ & \faThumbsUp$^{\ddagger}$ & \faThumbsUp$^{\ddagger}$ & \faThumbsUp$^{\ddagger}$ & \faThumbsUp$^{\ddagger}$ & \faThumbsUp$^{\ddagger}$ & \faThumbsUp & \noAnswer & $^{*}$can store vertices/edges in the same table.\\
Graph~\cite{ibm_db2_graph} & & & & & & & & & & & & & & & & & & & & $^{\ddagger}$inherited from the underlying IBM Db2™. \\
MS SQL Server & \faThumbsUp & \faThumbsUp & \faThumbsODown & \faThumbsODown & \faThumbsODown & \noAnswer & \noAnswer & \faThumbsODown & \noAnswer & \faThumbsODown& \faThumbsODown & \faThumbsUp & \faThumbsUp & \faThumbsUp & \faThumbsUp & \faThumbsUp & \faThumbsUp & \faThumbsUp & \faThumbsUp & The system uses an SQL graph extension. \\
2017~\cite{ms_sql_links} & & & & & & & & & & & & & & & & & & & & \\
OQGRAPH~\cite{oqgraph_links} & \faThumbsODown & \faThumbsUp & \faThumbsODown & \faThumbsODown & \faThumbsODown & \noAnswer$^{*}$ & \noAnswer$^{*}$ & \faThumbsODown & \faThumbsUp$^{*}$ & \faThumbsODown& \faThumbsODown & \faThumbsUp & \faThumbsUp & \faThumbsUp & \faThumbsUp & \faThumbsODown & \faThumbsUp & \faThumbsUp & \noAnswer & OQGRAPH uses MariaDB~\cite{bartholomew2012mariadb}.\\ \commentline{$^{*}$OQGRAPH uses row-oriented storage.}\\
SAP HANA~\cite{sap_hana_links} & \faThumbsUp & \faThumbsUp & \faThumbsODown & \faThumbsODown & \faThumbsODown & \faThumbsODown$^{*}$ & \faThumbsODown$^{*}$ & \faThumbsODown & \faThumbsODown$^{*}$ & \faThumbsODown & \faThumbsODown & \faThumbsUp & \faThumbsUp & \faThumbsUp & \faThumbsUp & \faThumbsUp & \faThumbsUp & \faThumbsUp & \faThumbsUp & $^{*}$SAP HANA is column-oriented, edges and\\
\commentline{vertices are stored in rows. SAP HANA can}\\
\commentline{be used with a dedicated graph engine~\cite{rudolf2013graph};}\\
\commentline{it offers some capabilities of a JSON document}\\
\commentline{store~\cite{sap_hana_links}}\\
SQLGraph~\cite{sqlgraph} & \faThumbsODown & \faThumbsUp & \faThumbsODown &  \faThumbsODown & \faThumbsODown & \noAnswer & \faThumbsUp$^{*}$ & \faThumbsODown & \faThumbsUp$^{\ddagger}$ & \faThumbsUp & \faThumbsODown & \faThumbsUp$^{\dagger}$ & \faThumbsUp$^{\dagger}$ & \faThumbsUp$^{\dagger}$ & \faThumbsUp$^{\dagger}$ & \faThumbsUp$^{\dagger}$ & \faThumbsUp$^{\dagger}$ & \faThumbsUp & \noAnswer & $^{*}$SQLGraph uses JSON for property storage. \\
\commentline{$^{\ddagger}$SQLGraph uses row-oriented storage.} \\
\commentline{$^{\dagger}$depends on the used SQL engine.} \\
\midrule
\multicolumn{21}{l}{\textbf{WIDE-COLUMN STORES} (\cref{widecolumn_section}). The main data model used: \textbf{key-value pairs} and \textbf{tables} (\cref{kv_section}, \cref{table_section}).} \\
\midrule
%
%
\textbf{JanusGraph~\cite{titan_link}} & \faThumbsODown & \faThumbsUp & \faThumbsODown & \faThumbsUp & \faThumbsODown & \faThumbsODown & \faThumbsUp & \faThumbsODown & \faThumbsODown & \faThumbsUp& \faThumbsODown & \faThumbsUp & \faThumbsUp & \faThumbsUp & \faThumbsUp & \faThumbsUp & \faThumbsUp & \faThumbsUp & \faThumbsUp & JanusGraph is the continuation of Titan.\\
\textbf{Titan~\cite{titan_link}} & \faThumbsUp & \faThumbsUp & \faThumbsODown & \faThumbsUp & \faThumbsODown & \faThumbsODown & \faThumbsUp & \faThumbsODown & \faThumbsODown & \faThumbsUp & \faThumbsODown & \faThumbsUp & \faThumbsUp & \faThumbsUp & \faThumbsUp & \faThumbsUp & \faThumbsUp & \faThumbsUp & \faThumbsUp & Enables various backends (e.g.,\\
\commentline{Cassandra~\cite{lakshman2010cassandra}).} \\
DSE Graph & \faThumbsODown & \faThumbsUp & \faThumbsODown & \faThumbsUp & \faThumbsODown & \faThumbsODown & \faThumbsUp & \faThumbsODown & \faThumbsODown & \faThumbsUp& \faThumbsODown & \faThumbsUp & \faThumbsUp & \faThumbsUp & \faThumbsUp & \noAnswer & \faThumbsOUp$^{*}$ & \faThumbsUp & \faThumbsUp & DSE Graph is based on Cassandra~\cite{lakshman2010cassandra}.\\
(DataStax) \cite{datastax_links} & & & & & & & & & & & & & & & & & & & & $^{*}$Support for AID, Consistency is configurable. \\
HGraphDB~\cite{hgraphdb_links} & \faThumbsODown & \faThumbsUp & \faThumbsODown & \faThumbsODown & \faThumbsODown & \faThumbsODown & \faThumbsUp & \faThumbsODown & \faThumbsUp & \faThumbsODown& \faThumbsODown & \faThumbsUp & \faThumbsUp & \faThumbsUp & \noAnswer & \noAnswer & \faThumbsODown$^{*}$ & \faThumbsUp & \faThumbsUp & HGraphDB uses TinkerPop3 with HBase~\cite{george2011hbase}.\\ \commentline{$^{*}$ACID is supported only within a row.} \\
\midrule
\multicolumn{21}{l}{\textbf{TUPLE STORES} (\cref{tuple_section}). The main data model used: \textbf{tuples} (\cref{tuple_model_section}).} \\
\midrule
\textbf{WhiteDB~\cite{whitedb_links}} & \faThumbsODown & \faThumbsODown & \faThumbsOUp$^{*}$ & \faThumbsOUp$^{\ddagger}$ & \faThumbsODown & \faThumbsODown & \faThumbsUp$^{*}$ & \faThumbsUp & \faThumbsOUp$^{\ddagger}$ & \faThumbsOUp$^{\ddagger}$& \faThumbsOUp$^{\ddagger}$ & \faThumbsODown & \faThumbsODown & \faThumbsODown & \faThumbsUp & \faThumbsUp & \faThumbsOUp$^{\dagger}$ & \faThumbsUp & \noAnswer &$^{*}$Implicit support for triples of integers.\\ \commentline{$^{\ddagger}$Implementable by the user. $^{\dagger}$Transactions}\\ \commentline{use a global shared/exclusive lock.} \\
Graphd~\cite{graphd_links} & \faThumbsODown & \faThumbsODown & \faThumbsOUp$^{*}$ & \faThumbsODown & \faThumbsODown & \noAnswer & \noAnswer & \noAnswer & \faThumbsUp & \faThumbsODown& \faThumbsODown & \faThumbsUp & \faThumbsUp & \noAnswer & \faThumbsUp & \noAnswer & \faThumbsOUp$^{\ddagger}$ & \noAnswer & \noAnswer & Backend of Google Freebase. \\ \commentline{$^{*}$Implicit support for triples. $^{\ddagger}$Subset of ACID.} \\
\midrule
\multicolumn{21}{l}{\textbf{OBJECT-ORIENTED DATABASES (OODBMS)} (\cref{objectoriented_section}). The main data model used: \textbf{objects} (\cref{object_section}).} \\
\midrule
\textbf{Velocity-} & \faThumbsUp & \faThumbsUp & \faThumbsODown & \faThumbsUp & \faThumbsODown & \faThumbsODown & \faThumbsUp & \faThumbsUp & \faThumbsUp & \faThumbsODown & \faThumbsODown & \faThumbsUp & \faThumbsUp & \faThumbsUp & \faThumbsUp & \faThumbsUp & \faThumbsUp & \faThumbsUp & \faThumbsUp & The system is based on VelocityDB~\cite{velocitydb_links}\\
\textbf{Graph~\cite{velocitygraph_links}} & & & & & & & & & & & & & & & & & & & & \\
Objectivity & \faThumbsODown & \faThumbsUp & \faThumbsODown & \noAnswer & \noAnswer & \noAnswer & \noAnswer & \noAnswer & \faThumbsUp & \noAnswer & \noAnswer & \faThumbsUp & \faThumbsUp & \faThumbsUp & \faThumbsUp & \faThumbsUp & \faThumbsUp & \faThumbsUp & \faThumbsUp & The system is based on ObjectivityDB~\cite{guzenda2000objectivity}.\\
ThingSpan~\cite{thingspan_links} & & & & & & & & & & & & & & & & & & & & \\
\midrule
\iftr
\multicolumn{21}{l}{\textbf{DATA HUBS} (\cref{datahub_section}). The main data model used: \textbf{several different ones}.} \\
\fi
\midrule
MarkLogic~\cite{marklogic_links} & \faThumbsUp & \faThumbsODown$^{*}$ & \faThumbsUp & \faThumbsODown & \faThumbsODown & \noAnswer & \faThumbsUp & \faThumbsODown & \faThumbsUp$^{*}$ & \faThumbsODown & \faThumbsODown & \faThumbsUp & \faThumbsUp & \faThumbsUp & \faThumbsUp & \noAnswer & \faThumbsUp & \faThumbsUp & \faThumbsUp & Supported storage/models: relational tables,\\
\commentline{RDF, various documents. $^{*}$Vertices are stored}\\
\commentline{as documents, edges are stored as RDF triples.}\\
OpenLink & \faThumbsUp & \faThumbsODown & \faThumbsUp & \faThumbsODown & \faThumbsODown & \noAnswer & \noAnswer & \faThumbsODown & \faThumbsODown & \faThumbsODown & \faThumbsODown & \faThumbsUp & \faThumbsUp & \faThumbsUp & \faThumbsUp & \faThumbsOUp$^{*}$ & \faThumbsUp & \faThumbsUp & \faThumbsUp & Supported storage/models: relational tables\\
Virtuoso~\cite{virtuoso_links} & & & & & & & & & & & & & & & & & & & & and RDF triples. $^{*}$This feature can be used\\
\commentline{relational data only.} \\
Cayley~\cite{cayley_links} & \faThumbsUp & \faThumbsUp & \faThumbsUp & \noAnswer & \noAnswer & \noAnswer & \faThumbsUp & \noAnswer & \noAnswer & \noAnswer& \noAnswer & \faThumbsUp & \faThumbsUp & \faThumbsODown & \faThumbsUp & \noAnswer & \faThumbsUp$^{*}$ & \noAnswer & \noAnswer & Supported storage/models: relational tables,\\
\commentline{RDF, document, key-value. $^{*}$This feature}\\
\commentline{depends on the backend.} \\
InfoGrid~\cite{infogrid_links} & \faThumbsUp & \faThumbsUp & \faThumbsODown & \noAnswer & \noAnswer & \noAnswer & \faThumbsUp & \noAnswer & \faThumbsUp & \faThumbsODown& \faThumbsODown & \faThumbsUp & \faThumbsUp & \faThumbsUp & \faThumbsUp & \noAnswer & \faThumbsOUp$^{*}$ & \noAnswer & \noAnswer & Supported storage/models: relational tables,\\
\commentline{Hadoop's filesystem, grid storage. $^{*}$A weaker}\\
\commentline{consistency model is used instead of ACID.} \\
Stardog~\cite{stardog_links} & \faThumbsOUp & \faThumbsUp$^{*}$ & \faThumbsOUp$^*$ & \faThumbsODown & \faThumbsODown & \noAnswer & \faThumbsUp & \faThumbsODown & \faThumbsUp$^{*}$ & \faThumbsODown & \faThumbsODown & \faThumbsUp & \faThumbsUp & \faThumbsODown & \faThumbsUp & \noAnswer & \faThumbsUp & \faThumbsUp & \faThumbsUp & Supported storage/models: relational tables,\\
\commentline{documents. $^{*}$RDF is simulated on relational}\\
\commentline{tables. Both LPG and RDF are enabled}\\
\commentline{through virtual quints.} \\
\bottomrule
\end{tabular}
\caption{
\textbf{Comparison of graph databases (RDBMS, wide-column stores, tuple stores, OODBMS, data hubs)}.
\textbf{\ul{Bolded systems}} are described in more detail in the corresponding
sections.
\iftr
\textbf{\ul{oB}}: A system supports secondary data models / backend types (in addition to its primary one).
\textbf{\ul{lpg}}, \textbf{\ul{rdf}}: A system supports, respectively, the
\textbf{Labeled Property Graph} and \textbf{RDF} without prior data
transformation.
\textbf{\ul{am}, \ul{al}}: The structure is represented as the \textbf{adjacency matrix} or the \textbf{adjacency list}.
\textbf{\ul{fs}}, \textbf{\ul{vs}}: Data records are \textbf{fixed size} and
\textbf{variable size}, respectively.
\textbf{\ul{dp}}: A system can use \textbf{direct
pointers} to link records. This enables {storing and traversing adjacency data without
maintaining indices}.
\textbf{\ul{se}}: Edges can be \textbf{stored in a separate edge record}.
\textbf{\ul{sv}}: Edges can be \textbf{stored in a vertex record}.
\textbf{\ul{lw}}: Edges can be \textbf{lightweight} (containing just a vertex ID or a pointer, both stored in a vertex record).
\textbf{\ul{ms}}: A system can operate in a \textbf{Multi Server} (distributed) mode.
\textbf{\ul{rp}}: Given a distributed mode, a system enables \textbf{Replication} of datasets.
\textbf{\ul{sh}}: Given a distributed mode, a system enables \textbf{Sharding} of datasets.
\textbf{\ul{ce}}: Given a distributed mode, a system enables \textbf{Concurrent Execution} of multiple queries.
\textbf{\ul{pe}}: Given a distributed mode, a system enables \textbf{Parallel Execution} of single queries on multiple nodes/CPUs.
\textbf{\ul{tr}}: Support for \textbf{ACID Transactions}.
\textbf{\ul{oltp}}: Support for \textbf{Online Transaction Processing}.
\textbf{\ul{olap}}: Support for \textbf{Online Analytical Processing}.
\faThumbsUp: A system offers a given feature.
\faThumbsOUp: A system offers a given feature in a limited way.
\faThumbsODown: A system does not offer a given feature.
\noAnswer: Unknown.\\
\fi
}
\label{survey-table2}
\end{table}

\begin{table}[hbtp]
%
%
\setlength{\tabcolsep}{1pt}
\renewcommand{\arraystretch}{0.7}
\centering
\ssmall
\sf
\begin{tabular}{llllllll}
\toprule
\multirow{2}{*}{\makecell[l]{\textbf{Graph Database} \textbf{System}}}
& \multicolumn{6}{c}{Graph Database Query Language}
& \multirow{2}{*}{\textbf{Other languages and additional remarks}} \\
\cmidrule(lr){2-7}
& \textbf{SPARQL} & \textbf{Gremlin} & \textbf{Cypher} & \textbf{SQL} & \textbf{GraphQL} & \textbf{Progr.~API} &  \\ 
\midrule
\multicolumn{8}{l}{\textbf{NATIVE GRAPH DATABASES (RDF model based, triple stores)} (\cref{triple_store_section}).} \\ 
\midrule 
AllegroGraph & \faThumbsUp & \faTimes & \faTimes & \faTimes & \faTimes & \faTimes & \faTimes \\ 
Amazon Neptune & \faThumbsUp & \faThumbsUp & \faTimes & \faTimes & \faTimes & \faTimes & \faTimes \\ 
AnzoGraph & \faThumbsUp & \faTimes & \faThumbsUp & \faTimes & \faTimes & \faTimes & \faTimes \\ 
Apache Jena TDB & \faThumbsUp & \faTimes & \faTimes & \faTimes & \faTimes & \faThumbsUp\ (Java) & \faTimes \\
Apache Marmotta & \faThumbsUp & \faTimes & \faTimes & \faTimes & \faTimes & \faTimes & Apache Marmotta also supports its native LDP and LDPath languages. \\
BlazeGraph & \faThumbsOUp$^*$ & \faThumbsUp & \faTimes & \faTimes & \faTimes & \faTimes & $^*$BlazeGraph offers SPARQL* to query RDF*. \\
BrightstarDB & \faThumbsUp & \faTimes & \faTimes & \faTimes & \faTimes & \faTimes & \faTimes \\ 
Cray Graph Engine & \faThumbsUp & \faTimes & \faTimes & \faTimes & \faTimes & \faTimes & \faTimes \\ 
gStore & \faThumbsUp & \faTimes & \faTimes & \faTimes & \faTimes & \faTimes & \faTimes \\
Ontotext GraphDB & \faThumbsUp & \faTimes & \faTimes & \faTimes & \faTimes & \faTimes & \faTimes \\ 
Profium Sense & \faThumbsUp & \faTimes & \faTimes & \faTimes & \faTimes & \faTimes & \faTimes \\ 
TripleBit & \faThumbsUp & \faTimes & \faTimes & \faTimes & \faTimes & \faTimes & \faTimes \\ 
\midrule
\multicolumn{8}{l}{\textbf{NATIVE GRAPH DATABASES (LPG model based)} (\cref{nativegdb_section}).} \\ 
\midrule
Gbase & \faTimes & \faTimes & \faTimes & \faThumbsUp & \faTimes & \faTimes & \faTimes \\ 
GraphBase & \faTimes & \faTimes & \faTimes & \faTimes & \faTimes & \faTimes & GraphBase uses its native query language. \\ 
Graphflow & \faTimes & \faTimes & \faThumbsOUp$^*$$^{\ddagger}$ & \faTimes & \faTimes & \faTimes & $^*$Graphflow supports a subset of Cypher~\cite{graphflow2019join}. $^{\ddagger}$Graphflow supports\\
& & & & & & & Cypher++ extension with subgraph-condition-action triggers~\cite{graphflow}. \\
LiveGraph & \faTimes & \faTimes & \faTimes & \faTimes & \faTimes & \faTimes & No focus on languages and queries. \\
Memgraph & \faTimes & \faTimes & \faThumbsUp$^*$ & \faTimes & \faTimes & \faTimes & $^*$openCypher. \\ 
Neo4j & \faTimes & \faThumbsOUp$^*$ & \faThumbsUp & \faTimes & \faThumbsUp$^{\ddagger}$ & \faThumbsOUp$^{\dagger}$ & $^*$Gremlin is supported as a part of TinkerPop integration.\\
& & & & & & & $^{\ddagger}$GraphQL supported with the GRANDstack layer.\\
& & & & & & & $^{\dagger}$Neo4j can be embedded in Java applications. \\
Sparksee/DEX & \faTimes & \faThumbsUp & \faTimes & \faTimes & \faTimes & \faThumbsUp\ (.NET)$^*$ & $^*$Sparksee/DEX also supports C++, Python, Objective-C, and Java APIs. \\
TigerGraph & \faTimes & \faTimes & \faTimes & \faTimes & \faTimes & \faTimes & TigerGraph uses GSQL~\cite{tiger_graph_links}. \\ 
Weaver & \faTimes & \faTimes & \faTimes & \faTimes & \faTimes & \faThumbsUp\ (C)$^*$ & $^*$Weaver also supports C++, Python. \\
\midrule
\multicolumn{8}{l}{\textbf{TUPLE STORES} (\cref{tuple_section}).} \\ 
\midrule
Graphd & \faTimes & \faTimes & \faTimes & \faTimes & \faTimes & \faTimes & Graphd uses MQL~\cite{graphd_links}. \\ 
WhiteDB & \faTimes & \faTimes & \faTimes & \faTimes & \faTimes & \faThumbsUp\ (C)$^*$ & $^*$WhiteDB also supports Python. \\
\midrule
\multicolumn{8}{l}{\textbf{DOCUMENT STORES} (\cref{document_store_section}).} \\ 
\midrule
ArangoDB & \faTimes & \faThumbsUp & \faTimes & \faTimes & \faTimes & \faTimes & ArangoDB uses AQL (ArangoDB Query Language). \\ 
Azure Cosmos DB & \faTimes & \faThumbsOUp & \faTimes & \faThumbsUp & \faTimes & \faTimes & \faTimes \\ 
Bitsy & \faTimes & \faThumbsUp & \faTimes & \faTimes & \faTimes & \faTimes & Bitsy also supports other Tinkerpop-compatible languages such as\\
& & & & & & & SQL2Gremlin and Pixy. \\
FaunaDB & \faTimes & \faTimes & \faTimes & \faTimes & \faThumbsUp & \faTimes & \faTimes \\ 
OrientDB & \faThumbsUp & \faThumbsUp & \faThumbsUp & \faThumbsUp$^*$ & \faTimes & \faThumbsUp\ (Java)$^{\ddagger}$ & $^*$An SQL extension for graph queries. $^{\ddagger}$OrientDB offers bindings to C,\\
& & & & & & & JavaScript, PHP, .NET, Python, and others. \\
\midrule
\multicolumn{8}{l}{\textbf{KEY-VALUE STORES} (\cref{keyvalue_section}).} \\ 
\midrule
Dgraph & \faTimes & \faTimes & \faTimes & \faTimes & \faThumbsUp$^*$ & \faTimes & $^*$A variant of GraphQL. \\ 
HyperGraphDB & \faTimes & \faTimes & \faTimes & \faTimes & \faTimes & \faThumbsUp\ (Java) & \faTimes \\
MS Graph Engine & \faTimes & \faTimes & \faTimes & \faTimes & \faTimes & \faTimes & MS Graph Engine uses LINQ~\cite{trinity_paper}. \\
RedisGraph & \faTimes & \faTimes & \faThumbsUp & \faTimes & \faTimes & \faTimes & \faTimes \\ 
\midrule
\multicolumn{8}{l}{\textbf{WIDE-COLUMN STORES} (\cref{widecolumn_section}).} \\ 
\midrule
DSE Graph (DataStax) & \faTimes & \faThumbsUp & \faTimes & \faTimes & \faTimes & \faTimes & DSE Graph also supports CQL~\cite{datastax_links}. \\ 
HGraphDB & \faTimes & \faThumbsUp & \faTimes & \faTimes & \faTimes & \faTimes & \faTimes \\ 
JanusGraph & \faTimes & \faThumbsUp & \faTimes & \faTimes & \faTimes & \faTimes & \faTimes \\
Titan & \faTimes & \faThumbsUp & \faTimes & \faTimes & \faTimes & \faTimes & \faTimes \\
\midrule
\multicolumn{8}{l}{\textbf{RELATIONAL DBMS (RDBMS)} (\cref{rdbms_section}).} \\
\midrule
AgensGraph & \faTimes & \faTimes & \faThumbsUp$^*$ & \faThumbsUp$^{\ddagger}$ & \faTimes & \faTimes & $^*$A variant called openCypher~\cite{green2018opencypher, marton2017formalising}. $^{\ddagger}$ANSI-SQL. \\ 
FlockDB & \faTimes & \faTimes & \faTimes & \faThumbsUp & \faTimes & \faTimes & FlockDB uses the Gizzard framework and MySQL. \\
IBM Db2 Graph & \faTimes & \faThumbsOUp$^*$ & \faTimes & \faThumbsUp & \faTimes & \faThumbsUp\ (Java)$^{\ddagger}$ & $^*$IBM Db2 Graph supports only graph queries whose results can be\\
& & & & & & & returned to rows. $^{\ddagger}$IBM Db2 Graph also supports Scala, Python and\\
& & & & & & & Groovy. \\
MS SQL Server 2017 & \faTimes & \faTimes & \faTimes & \faThumbsUp$^*$ & \faTimes & \faTimes & $^*$Transact-SQL. \\
OQGRAPH & \faTimes & \faTimes & \faTimes & \faThumbsUp & \faTimes & \faTimes & \faTimes \\ 
Oracle Spatial and Graph & \faThumbsUp & \faTimes & \faTimes & \faThumbsUp$^*$ & \faTimes & \faTimes & $^*$PGQL~\cite{van2016pgql}, an SQL-like graph query language. \\
SAP HANA & \faTimes & \faTimes & \faTimes & \faThumbsUp$^{*}$ & \faTimes & \faThumbsUp$^{\ddagger}$ & $^*$SAP HANA offers bindings to Rust, ODBC, and others.\\
& & & & & & & $^{\ddagger}$GraphScript, a domain-specific graph query language. \\
SQLGraph & \faTimes & \faThumbsOUp$^*$ & \faTimes & \faThumbsOUp$^{\ddagger}$  & \faTimes & \faTimes & $^*$SQLGraph doesn't support Gremlin side effect pipes.\\
& & & & & & & $^{\ddagger}$Graph is encoded in a way specific to SQLGraph. \\
\midrule
\multicolumn{8}{l}{\textbf{OBJECT-ORIENTED DATABASES (OODBMS)} (\cref{objectoriented_section}).} \\ 
\midrule
Objectivity ThingSpan & \faTimes & \faTimes & \faTimes & \faTimes & \faTimes & \faTimes & Objectivity ThingSpan uses a native DO query language~\cite{thingspan_links}. \\ 
VelocityGraph & \faTimes & \faTimes & \faTimes & \faTimes & \faTimes & \faThumbsUp\ (.NET) & \faTimes \\
\midrule
\iftr
\multicolumn{8}{l}{\textbf{DATA HUBS} (\cref{datahub_section}).} \\ 
\fi
\midrule
Cayley & \faTimes & \faThumbsUp$^*$ & \faTimes & \faTimes & \faThumbsUp & \faTimes & $^*$Cayley supports Gizmo, a Gremlin dialect~\cite{cayley_links}.\\
& & & & & & & Cayley also uses MQL~\cite{cayley_links}. \\
InfoGrid & \faTimes & \faTimes & \faTimes & \faTimes & \faTimes & \faThumbsUp\ (REST) & \faTimes \\
MarkLogic & \faTimes & \faTimes & \faTimes & \faTimes & \faTimes & \faTimes & MarkLogic uses XQuery~\cite{boag2002xquery}. \\ 
OpenLink Virtuoso & \faThumbsUp & \faTimes & \faTimes & \faThumbsUp & \faTimes & \faTimes & OpenLink Virtuoso also supports XQuery~\cite{boag2002xquery}, XPath v1.0~\cite{clark1999xml},\\
& & & & & & & and XSLT v1.0~\cite{kay2001xslt}. \\
Stardog & \faThumbsUp$^*$ & \faThumbsUp & \faTimes & \faTimes & \faThumbsUp & \faTimes & $^*$Stardog supports the Path Query extension~\cite{stardog_links}. \\
\bottomrule
\end{tabular}
\caption{
\textbf{Support for different graph database query languages in different graph database systems}.
``\textbf{Progr.~API}'' determines whether a given system supports formulating queries using some native programming language such as C++.
``\faThumbsUp'': A system supports a given language.
``\faThumbsOUp'': A system supports a given language in a limited way.
``\faTimes'': A system does not support a given language.\\
}
\label{tab:languages}
%
%
\end{table}

\section{ANALYSIS OF DATABASE SYSTEMS}
\label{existing_gdb}

We survey and describe selected graph database systems with respect to the
proposed taxonomy. In each system category, we describe selected
representative systems, focusing on the associated graph model, as well as
data and storage organization.
Tables~\ref{survey-table1} and~\ref{survey-table2} illustrate the details of
different graph database systems, including the ones described in this section\footnote{We encourage participation
in this survey. In case the reader is
in possession of additional information relevant for the tables,
the authors would welcome the input.}.
\iftr
The tables indicate which features are supported by
which systems. We use symbols ``\faThumbsUp'', ``\faThumbsOUp'', and
``\faThumbsODown'' to indicate that a given system offers a
given feature, offers a given feature in a limited way, and does not offer a
given feature, respectively.  
\fi
``\noAnswer'' indicates we were unable to infer this
information based on the available documentation.
We report the support for different graph query languages in
Table~\ref{tab:languages}.
Finally, we analyze different taxonomy aspects
in~\cref{sec:main-discussion}.

\subsection{Discussion on Selection Criteria}

When selecting systems for consideration in the survey, we use two
criteria. First, we use the DB-Engines Ranking\footnote{\url{https://db-engines.com/en/ranking/graph+dbms}} to select the most
popular systems in each considered backend category. We
also pick interesting research systems (e.g, SQLGraph~\cite{sqlgraph},
LiveGraph~\cite{livegraph}, or Weaver~\cite{dubey2016weaver}) which
are not included in this ranking.
For detailed discussions, we also consider the availability of technical
details (i.e., most systems are closed source or do not offer any
design details).

\subsection{RDF Stores (Triple Stores)}
\label{triple_store_section}

RDF stores~\cite{broekstra2002sesame, abadi2007scalable,
neumann2010x, harth2007yars2}, also called triple stores, implement the
Resource Description Framework (RDF) model (\cref{rdf_section}). These systems
organize data into triples. We now describe in more detail a selected recent
RDF store, Cray Graph Engine (\cref{cge_section}). We also provide more details
on two other systems, AllegroGraph and BlazeGraph, focusing on \emph{variants
of the RDF model} used in these systems (\cref{ab_section}).

\subsubsection{Cray Graph Engine}
\label{cge_section}

Cray Graph Engine (CGE)~\cite{cge_paper}
is a triple store that can scale to a trillion RDF
triples.
CGE does not store triples but \textit{quads} (4-tuples), where the
fourth element is a graph ID. Thus, one can store multiple graphs in
one CGE database.
Quads in CGE are grouped by their predicate and the identifier of the graph
that they are a part of. Thus, only a pair with a subject and an object needs
to be stored for one such group of quads. These subject/object pairs are stored
in hashtables (one hashtable per group). Since each subject and object is
represented as a unique 48-bit integer identifier (HURI), the subject/object
pairs can be packed into 12 bytes and stored in a 32-bit unsigned integer
array, ultimately reducing the amount of needed storage.
%

\subsubsection{AllegroGraph and BlazeGraph}
\label{ab_section}

There exist many other RDF graph databases. We briefly describe two systems
that extend the original RDF model: AllegroGraph and BlazeGraph.

First, some RDF stores allow for attaching \emph{attributes} to a triple
explicitly. AllegroGraph~\cite{allegro_graph_links} allows an arbitrary set of
attributes to be defined per triple when the triple is created. However, these
attributes are immutable. Figure~\ref{fig:rdf:transformation_allegrodb}
presents an example RDF graph with such attributes.
%
%
This figure uses the same LPG graph as in previous examples provided  in
Figure~\ref{fig:rdf:transformation_1} and
Figure~\ref{fig:rdf:transformation_2}, which contain example transformations
from the LPG into the original RDF model.

\begin{figure}[h]
	\centering
	\includegraphics[width=0.9\textwidth]{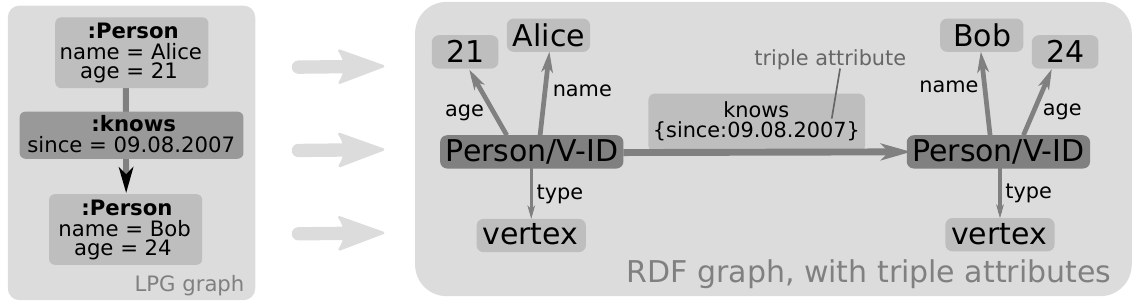}
	\caption{\textbf{Comparison of an LPG graph and an RDF graph:} a
    transformation from LPG to RDF \textbf{with triple attributes}. We represent the
    triple attributes as a set of key-value pairs.
    ``Person/V-ID'', ``age'', ``name'', ``type'' and ``knows'' are RDF URIs.
    The transformation uses the assumption that there is one label per vertex and edge.
    }
	\label{fig:rdf:transformation_allegrodb}
%
%
\end{figure}

Second, BlazeGraph~\cite{blaze_graph_links} implements
\textit{RDF*}~\cite{rdfstar_link, hartig2017rdf}, an augmentation of RDF that allows for
attaching triples to triple predicates (see Figure~\ref{fig:rdf:transformation_star}).
Vertices can use triples for storing labels and properties, analogously as with
the plain RDF. However, with RDF*, one can represent LPG edges more naturally
than in the plain RDF. Specifically, edges can be stored as triples, and edge
properties can be linked to the edge triple via other triples.

\begin{figure}[h]
	\centering
	\includegraphics[width=0.9\textwidth]{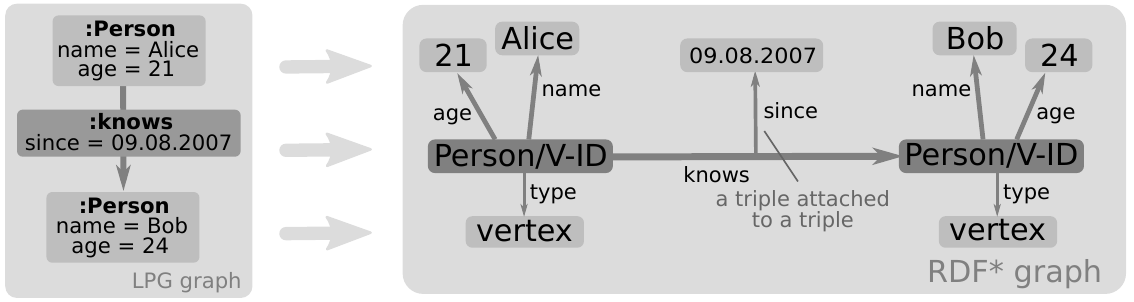}
\caption{\textbf{Comparison of an LPG graph and an RDF* graph:} a
    transformation from LPG to RDF*, that enables \textbf{attaching triples to triple predicates}.
    ``Person/V-ID'', ``age'', ``name'', ``type'', ``since'' and ``knows'' are RDF URIs.
    The transformation uses the assumption that there is one label per vertex and edge.
    }
	\label{fig:rdf:transformation_star}
%
%
\end{figure}

\subsection{Tuple Stores}
\label{tuple_section}

A tuple store is a generalization of an RDF store. RDF
stores are restricted to triples (or quads, as in CGE) whereas tuple stores can
maintain tuples of arbitrary sizes, as detailed
in~\cref{tuple_model_section}.
%

\subsubsection{WhiteDB}
\label{whitedb_section}

WhiteDB~\cite{whitedb_links} is a tuple store that enables allocating new records (tuples) with an arbitrary tuple
length (number of tuple elements). Small values and pointers to other tuples
are stored directly in a given field. Large strings are kept in a separate
store. Each large value is only stored once, and a reference counter keeps
track of how many tuples refer to it at any time. 
WhiteDB only enables accessing single tuple records, there is no
higher level query engine or graph API that would allow to, for example,
execute a query that fetches all neighbors of a given vertex.  However, one can
use tuples as vertex and edge storage, linking them to one another via
memory pointers.
This facilitates fast resolution of various queries about the structure of
an arbitrary irregular graph structure in WhiteDB. For example, one can store a
vertex~$v$ with its properties as consecutive fields in a tuple
associated with~$v$, and maintain pointers to selected neighborhoods
of~$v$ in $v$'s tuple. More examples on using WhiteDB (and
other tuple stores such as Graphd) for maintaining graph data can be found
online~\cite{whitedb_links, meyer2010optimizing}.

\subsection{Key-Value Stores}
\label{keyvalue_section}

One can also explicitly use key-value (KV) stores for maintaining a graph (cf.~\cref{kv_section}). 
\iftr
We
provide details of using a collection of key-value pairs to model a graph
in~\cref{kv_section}. Here, we describe {selected} KV stores used as graph
databases: MS Graph Engine (also called Trinity) and HyperGraphDB.
\fi

\subsubsection{Microsoft's Graph Engine (Trinity)}
\label{ge_section}

Microsoft's Graph Engine~\cite{trinity_paper} is based on a
distributed KV store called Trinity. Trinity implements a globally addressable
distributed RAM storage. 
In Trinity, keys are called \textit{cell IDs} and values are called
\textit{cells}. A cell can hold data items of different data types, including
IDs of other cells.  MS Graph Engine introduces a graph storage layer on top of
the Trinity KV storage layer.  Vertices are stored in {cells}, where a
dedicated field contains a vertex ID or a hash of this ID. Edges adjacent to a
given vertex~$v$ are stored as a list of IDs of $v$'s neighboring
vertices, directly in $v$'s cell. However, if an edge holds rich data,
such an edge (together with the associated data) can also be stored in a
separate dedicated cell.

\subsubsection{HyperGraphDB}
\label{hyperGDB_section}

HyperGraphDB~\cite{hypergraphdb_links} stores hypergraphs (definition in~\cref{sec:hg_model}).
The basic building blocks of HyperGraphDB are \textit{atoms}, the values of the
KV store.  Every \textit{atom} has a cryptographically strong ID. This reduces
a chance of collisions (i.e., creating identical IDs for different graph
elements by different peers in a distributed environment). Both hypergraph
vertices and hyperedges are atoms. Thus, they have their own unique IDs.
An atom of a hyperedge stores a list of IDs corresponding to the vertices
connected by this hyperedge. Vertices and hyperedges also have a \emph{type
ID} (i.e., a label ID) and they can store additional data (such as
properties) in a recursive structure (referenced by a \emph{value ID}).  This
recursive structure contains value IDs identifying other atoms (with other
recursive structures) or binary data.
Figure~\ref{fig:hypergraphdb:mappings} shows an example of how a KV store is
used to represent a hypergraph in HyperGraphDB.

\begin{figure}[h]
	\centering
	\includegraphics[width=0.65\textwidth]{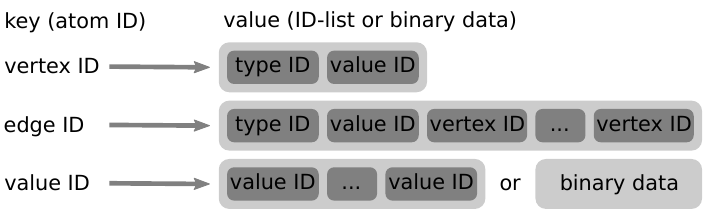}
	\caption{\textbf{An example utilization of key-value stores for maintaining hypergraphs in HyperGraphDB (a type is a term used in HyperGraphDB to refer to a label).}
	}
	\label{fig:hypergraphdb:mappings}
\end{figure}

\subsection{Document Stores}
\label{document_store_section}

In document stores, a fundamental storage unit is a document, described
in~\cref{doc_section}. We select two document stores for a more detailed
discussion, OrientDB and ArangoDB.

\subsubsection{OrientDB}
\label{orientdb_section}

In OrientDB~\cite{orientdb_links}, every document~$d$ has a \emph{Record ID
(RID)}, consisting of the ID of the \emph{collection of documents} where $d$ is
stored, and the \emph{position} (also referred to as the \emph{offset}) within
this collection. Pointers (called \emph{links}) between
documents are represented using these unique RIDs.

OrientDB~\cite{orientdb_links} introduces \textit{regular edges} and
\textit{lightweight edges}. {Regular edges} are stored in an
\emph{edge document} and can have their own associated key-value pairs (e.g.,
to encode edge properties or labels). {Lightweight edges}, on the other hand,
are stored directly in the document of the adjacent (source or destination)
vertex. Such edges do not have any associated key-value pairs. They constitute
simple pointers to other vertices, and they are implemented as document RIDs.
Thus, a vertex document not only stores the labels and properties of the
vertex, but also a list of lightweight edges (as a list of RIDs of the
documents associated with neighboring vertices), and a list of pointers to the
adjacent regular edges (as a list of RIDs of the documents associated with
these regular edges).  Each regular edge has pointers (RIDs) to the documents
storing the source and the destination vertex.  Each vertex stores a list of
links (RIDs) to its incoming and the outgoing edges.

Figure~\ref{fig:orientdb:storage_overview} contains an example of using
documents for representing vertices, regular edges, and lightweight edges in
OrientDB. Figure~\ref{fig:orientdb:storage_documents} shows example vertex and edge
documents.

  
\begin{figure}[h]
\centering
\includegraphics[width=0.5\textwidth]{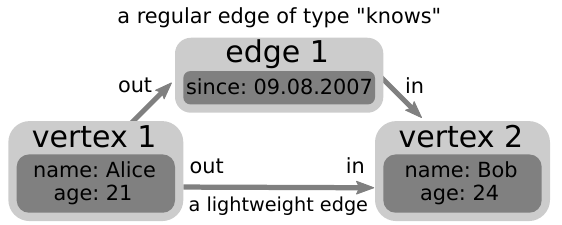}
\caption{Two vertex documents connected with \textbf{a lightweight edge and a
regular edge (knows) in OrientDB}.}
\label{fig:orientdb:storage_overview}
%
\end{figure}

\begin{figure}[h]
\centering
\iftr
\includegraphics[width=0.95\textwidth]{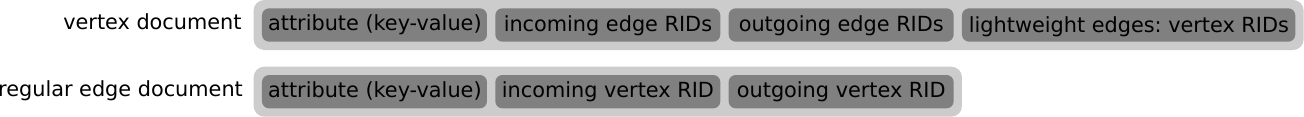}
\fi
\caption{Example OrientDB \textbf{vertex and edge documents (complex JSON documents are also supported)}.}
\label{fig:orientdb:storage_documents}
%
\end{figure}

\subsubsection{ArangoDB}
\label{arangodb_design}

ArangoDB~\cite{arangodb_links, arangodb_indexing_links} keeps its documents in
a \emph{binary format} called \emph{VelocyPack}, which is a compacted
implementation of JSON documents. Documents can be stored in different
collections and have a \textit{\_key} attribute which is a unique ID within a
given collection. Unlike OrientDB, these IDs are no direct memory pointers.
For maintaining graphs, ArangoDB uses \emph{vertex collections} and \emph{edge
collections}. The former are regular document collections with vertex
documents. Vertex documents store no information about adjacent edges.
This has the advantage that a vertex document does not
have to be modified when one adds or removes edges. Second, edge collections
store edge documents. Edge documents have two particular properties:
\textit{\_from} and \textit{\_to}, which are the IDs of the documents
associated with two vertices connected by a given edge.
An optimization in ArangoDB's design prevents reading vertex documents and
enables directly accessing one edge document based on the vertex ID
\emph{within another edge document}. This may improve cache efficiency and thus
reduce query execution time~\cite{arangodb_indexing_links}.

One can use
different collections of documents to store different edge types (e.g.,
``friend\_of'' or ``likes'').  When retrieving edges conditioned on some edge
type (e.g., ``friend\_of''), one does not have to traverse the whole adjacency
list (all ``friend\_of'' and ``likes'' edges).  Instead, one can target the
collection with the edges of the specific edge type (``friend\_of'').

\subsection{Wide-Column Stores}
\label{widecolumn_section}

Wide-column stores combine different features of key-value stores and
relational tables. On one hand, a wide-column store maps keys to \textit{rows}
(a KV store that maps keys to values). Every {row} can have an arbitrary number
of \textit{cells} and every cell constitutes a key-value pair.  Thus, a {row}
contains a mapping of cell keys to cell values, effectively making a
wide-column store a \emph{two-dimensional KV store} (a row key and a cell key
both identify a specific value). On the other hand, a wide-column store is a
\textit{table}, where cell keys constitute column names. However, unlike in a
relational database, the names and the format of columns may differ between
rows within the same table. We illustrate an example subset of rows and
cells in a wide-column store in Figure~\ref{fig:titan:widecolumn}.

\begin{figure}[h]
	\centering
	\includegraphics[width=0.7\textwidth]{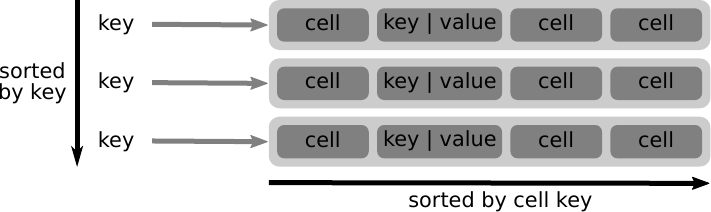}
	\caption{\textbf{An illustration of wide-column stores}: mapping keys to rows and column-keys to cells within the rows.}
	\label{fig:titan:widecolumn}
\end{figure}

\subsubsection{Titan and JanusGraph}
\label{titan_section}

Titan~\cite{titan_link} and its continuation
JanusGraph~\cite{janus_graph_links} are
built on top of wide-column stores. They can use different
wide-column stores as backends, for example Apache
Cassandra~\cite{cassandra_links}.
In both systems, when storing a graph, each row represents a vertex.  Each
vertex property and adjacent edge is stored in a separate cell.  One edge is
thus encoded in a single cell, including all the properties of this edge.
Since cells in each row are sorted by the cell key, this sorting order can be
used to find cells efficiently.  For graphs, cell keys for properties and edges
are chosen such that after sorting the cells, the cells storing properties come
first, followed by the cells containing edges, see
Figure~\ref{fig:titan:bigtable}. Since rows are ordered by the key, both
systems straightforwardly partition tables into so called \textit{tablets},
which can then be distributed over multiple servers.

\begin{figure}[h]
  \centering
	\includegraphics[width=0.8\textwidth]{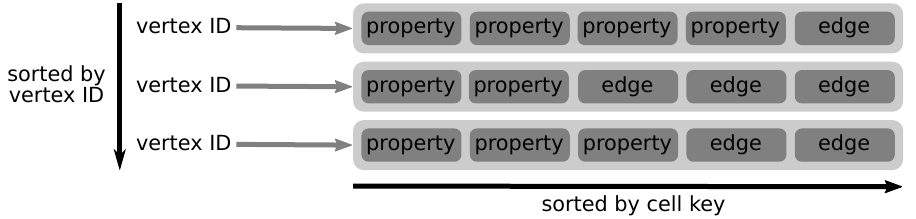}
	\caption{\textbf{An illustration of Titan and JanusGraph:} using wide-column stores for storing graphs. The illustration
  is inspired by and adapted from the work by Sharma~\cite{sharma2014cassandra}.}
	\label{fig:titan:bigtable}
\end{figure}

\subsection{Relational Database Management Systems}
\label{rdbms_section}

Relational Database Management Systems (RDBMS) store data in two dimensional
\emph{tables} with \emph{rows} and \emph{columns}, described in more detail in
the corresponding data model section in~\cref{table_section}.

There are two types of RDBMS: \emph{column} RDBMS (not to be confused with
\emph{wide-column} stores) and \emph{row} RDBMS (also referred to as
\emph{column-oriented} or \emph{columnar} and \emph{row-oriented}). They differ
in physical data persistence. In many row RDBMS, data
items in memory are kept in contiguous
rows~\cite{roozkhosh2021relational, redshift-columnar}. Column RDBMS,
on the other hand, store table columns contiguously. Row RDBMS are more
efficient when only a few rows need to be retrieved, but with
all their columns. Conversely, column RDBMS are more efficient when many rows
need to be retrieved, but only with a few columns. Graph database solutions
that use RDBMS as their backends use both row RDBMS (e.g., Oracle Spatial and
Graph~\cite{oracle_spatial}, OQGRAPH built on MariaDB~\cite{oqgraph_links}) and
column RDBMS (e.g., SAP HANA~\cite{sap_hana_links}).

\subsubsection{Oracle Spatial and Graph}
\label{oraclespatial_section}
 
Oracle Spatial and Graph~\cite{oracle_spatial} is
built on top of Oracle Database. It provides a rich set of tools for
administration and analysis of graph data. Oracle Spatial and Graph comes with
a range of built-in parallel graph algorithms (e.g., for community detection,
path finding, traversals, link prediction, PageRank, etc.). Both LPG and RDF
models are supported. Rows of RDBMS tables
constitute vertices and relationships between these rows form edges. Associated
properties and attributes are stored as key-value pairs in separate structures.

\subsection{Object-Oriented Databases}
\label{objectoriented_section}

Object-oriented database management systems (OODBMS)~\cite{oodbms_paper} enable
modeling, storing, and managing data in the form of \emph{language objects}
used in object-oriented programming languages.  We summarize such objects
in~\cref{object_section}.

\subsubsection{VelocityGraph}
\label{velocitygraph_section}

VelocityGraph~\cite{velocitygraph_links} is a graph database relying on the
VelocityDB~\cite{velocitydb_links} distributed object database. 
VelocityGraph edges, vertices, as well as edge or vertex properties are stored
in C\# objects that contain references to other objects. To handle this
structure, VelocityGraph introduces abstractions such as VertexType, EdgeType,
and PropertyType. Each object has a unique object identifier (Oid), pointing to
its location in physical storage. Each vertex and edge has one type (label).
Properties are stored in dictionaries. Vertices keep the adjacent edges
in collections.

\subsection{LPG-Based Native Graph Databases}
\label{nativegdb_section}

Graph database systems described in the previous sections with the exception of triple stores are all based on some
database backend that was not originally built just for managing graphs.  In
what follows, we describe \emph{LPG-based native graph databases}: systems that were
specifically build to maintain and process graphs.

\subsubsection{Neo4j: Direct Pointers}
\label{neo4j_design}

Neo4j~\cite{neo4j_book} is the most popular graph database system, according to
different database rankings (see the links provided in the introduction).
Neo4j implements the LPG model using a storage design based on fixed-size
records. A vertex~$v$ is represented with a \emph{vertex record}, which
stores (1) $v$'s labels, (2) a pointer to a linked list of $v$'s properties,
(3) a pointer to the first edge adjacent to $v$, and (4) some flags.  An
edge~$e$ is represented with an \emph{edge record}, which stores (1) $e$'s edge
type (a label), (2) a pointer to a linked list of $e$'s properties, (3) a
pointer to two vertex records that represent vertices adjacent to $e$, (4)
pointers to the ALs of both adjacent vertices, and (5) some flags.
Each property record can store up to four properties, depending on the size of
the property value. Large values (e.g., long strings) are stored in a separate
\emph{dynamic store}.
Storing properties outside vertex and edge records allows those records to be
small. Moreover, if no properties are accessed in a query, they are not 
loaded at all.
The AL of a vertex is implemented as a doubly linked list. An 
edge is stored once, but is part of two such linked lists (one list for each adjacent vertex).
Thus, an edge has two pointers to the previous edges and two pointers to the
next edges.
%
%
Figure~\ref{fig:neo4j:storage_overview} outlines the Neo4j design;
Figure~\ref{fig:neo4j:storage_records} shows the details of vertex and edge
records.

\begin{figure}[h]
\centering
\iftr
\includegraphics[width=0.6\textwidth]{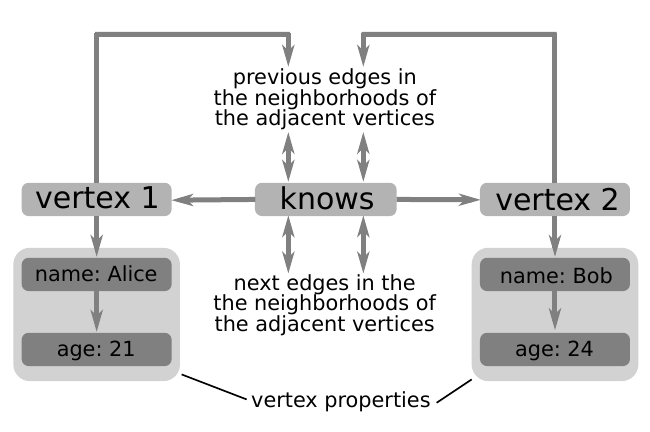}
\fi
\caption{\textbf{Summary of the Neo4j structure:} two vertices linked by a
``knows'' edge. Both vertices maintain linked lists of properties. The
edges are part of two doubly linked lists, one linked list per
adjacent vertex.}
\label{fig:neo4j:storage_overview}
%
\end{figure}

\begin{figure}[h]
\centering
\includegraphics[width=1.0\textwidth]{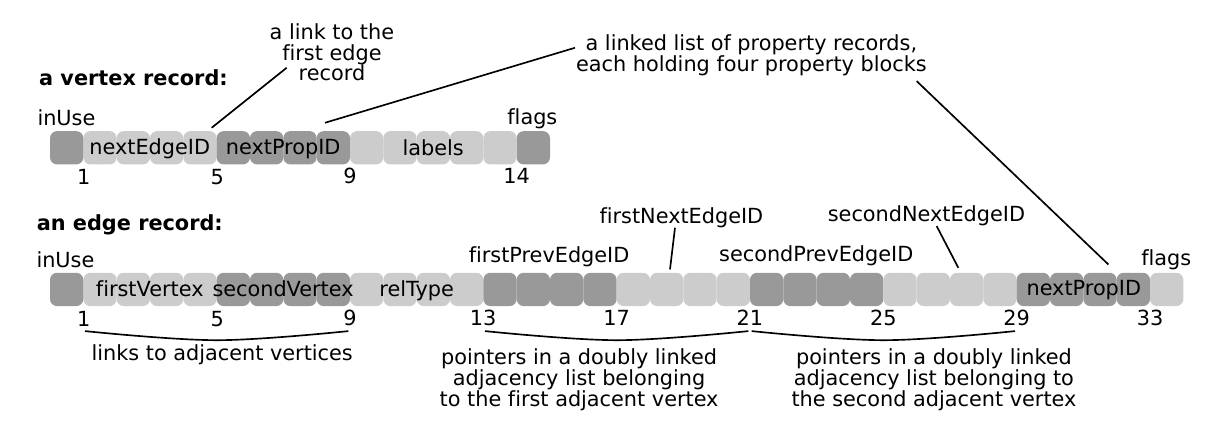}
\caption{\textbf{An overview of the Neo4j vertex and edge records.}}
\label{fig:neo4j:storage_records}
%
\end{figure}

A core concept in Neo4j is using \textit{direct pointers}~\cite{neo4j_book}: a
vertex stores pointers to the physical locations of its neighbors. Thus, for
neighborhood queries or traversals, one needs no index and can instead follow
direct pointers (except for the root vertices in traversals). Consequently, the
query complexity does not dependent on the graph size. Instead, it only depends
on how large the visited subgraph is\footnote{That said, if the graph does not
fit into the main memory, the execution speed heavily depends on caching and
cache pre-warming, i.e., the running time may significantly increase.}.

\subsubsection{Sparksee/DEX: B+ Trees and Bitmaps}
\label{dex_section}

Sparksee is a graph database system that was formerly known as
DEX~\cite{sparksee_paper}. Sparksee implements the LPG model in the following
way. Vertices and edges (both are called objects) are identified by unique
IDs. For each property name, there is an associated B+ tree that maps vertex
and edge IDs to the respective property values. The reverse mapping from a
property value to vertex and edge IDs is maintained by a bitmap, where a bit
set to one indicates that the corresponding ID has some property value. Labels
and vertices and edges are mapped to each other in a similar way.
Moreover, for each vertex, two bitmaps are stored: One bitmap indicates the
incoming edges, and another one the outgoing edges.
Furthermore, two B+ trees maintain the information about what vertices
an edge is connected to (one tree for each edge direction).
Figure~\ref{fig:sparksee:maps} illustrates example mappings.

\begin{figure}[h]
\centering
\includegraphics[width=0.7\textwidth]{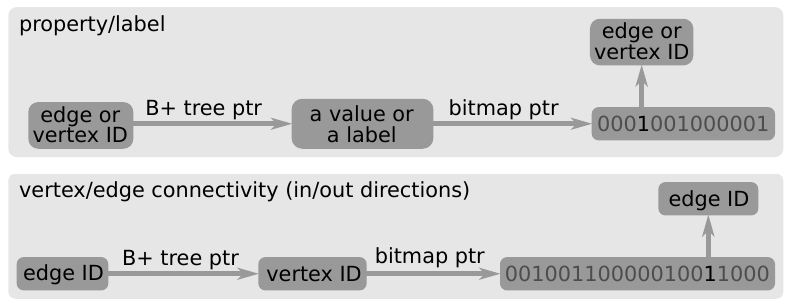}
\caption{\textbf{Sparksee maps for properties, labels, and
vertex/edge connectivity}. All mappings are bidirectional.}
\label{fig:sparksee:maps}
%
\end{figure}

Sparksee is one of the few systems that are \emph{not} record based. Instead,
Sparksee uses \emph{maps} implemented as B+ trees~\cite{b_tree_paper} and
bitmaps.
The use of bitmaps allows for some operations to be performed as bit-level
operations. For example, if one wants to find all vertices with certain values
of properties such as ``age'' and ``first name'', one can simply find two
bitmaps associated with the ``age'' and the ``first name'' properties, and then
derive a third bitmap that is a result of applying a bitwise AND operation to
the two input bitmaps.

Uncompressed bitmaps could grow unmanageably in size.  As most graphs are
sparse, bitmaps indexed by vertices or edges mostly contain zeros.  To
alleviate large sizes of such sparse bitmaps, they are cut into 32-bit
\textit{clusters}.  If a cluster contains a non-zero bit, it is stored
explicitly.  The bitmap is then represented by a collection of
(\textit{cluster-id}, \textit{bit-data}) pairs.  These pairs are stored in a
sorted tree structure to allow for efficient lookup, insertion, and deletion.

\iftr

\subsubsection{GBase: Sparse Adjacency Matrix Format}
\label{gbase_section}

GBase~\cite{gbase_paper} is a system that can only represent the
\emph{structure} of a directed graph; it stores neither properties nor labels.
The goal of GBase is to maintain a compression of the adjacency matrix of a
graph such that one can efficiently retrieval all incoming and outgoing edges
of a selected vertex without the prohibitive $O(n^2)$ matrix storage overheads.
Simultaneously, using the adjacency matrix enables verifying in $O(1)$ time
whether two arbitrary vertices are connected. 
%
%
To compress the adjacency matrix, GBase cuts it into $K^2$ quadratic blocks
(there are $K$ blocks along each row and column).  Thus, queries that fetch in-
and out-neighbors of each vertex require only to fetch $K$ blocks.  The
parameter $K$ can be optimized for specific databases.  When $K$ becomes
smaller, one has to retrieve more small files (assuming one block is stored in
one file).  If $K$ grows larger, there are fewer files but they become larger,
generating overheads.
Further optimizations can be made when blocks contain either only zeroes or
only ones; this enables higher compression rates.

\fi

\iftr

\subsection{Data Hubs}
\label{datahub_section}

Data hubs are systems that enable using multiple data models and corresponding
storage designs. They often combine relational databases with RDF, document,
and key-value stores. This can be beneficial for applications that require a
variety of data models, because it provides a variety of storage options in a
single unified database management system. One can keep using RDBMS features,
upon which many companies heavily rely, while also storing graph data.

\subsubsection{OpenLink Virtuoso}
\label{virtuoso_section}

OpenLink Virtuoso~\cite{virtuoso_links} provides RDBMS, RDF, and document
capabilities by connecting to a variety of storage systems.  Graphs are stored
in the RDF format only, thus the whole discussion from~\cref{rdf_section} also
applies to Virtuoso RDF.

\subsubsection{MarkLogic}
\label{marklogic_section}

MarkLogic~\cite{marklogic_links} models graphs with documents for vertices,
therefore allowing an arbitrary number of properties for vertices.
However, it uses RDF triples for edges.

\fi

\section{TAKEAWAYS, INSIGHTS, FUTURE DIRECTIONS}
\label{sec:main-discussion}

We now offer different insights about the described systems, considering both
practitioners and researchers. We interleave these insights
with suggestions for future developments and research.

\subsection{Discussion, Takeaways, and Insights on Data Organization}
\label{sec:discussion_data-org}

We discuss the data organization aspects of our taxonomy with respect
to specific graph databases.
\iftr
For a detailed description and analysis of all the considered
aspects, see~\cref{sec:taxo} and Tables~\ref{survey-table1}
and~\ref{survey-table2}.
\fi

\subsubsection{Conceptual Graph Models}

There is no one standard conceptual graph model, but two models have
proven to be popular: RDF and LPG. RDF is a well-defined standard. However, it
only supports simple triples (subject, predicate, object) representing edges
from subject identifiers via predicates to objects. LPG allows vertices and
edges to have labels and properties, thus enabling more natural data modeling in
different scenarios~\cite{neo4j_book}.
Still, it is not standardized, and there are many variants
(cf.~\cref{lpgvariants_section}); However, it is becoming standardized in the
upcoming SQL/PGQ and GQL standards from ISO~\cite{francis2023researcher}. Some systems limit the number of
labels to just one. For example, MarkLogic allows properties for
vertices but none for edges, and thus can be viewed as a combination of LPG
(vertices) and RDF (edges).
Data stored in the LPG model can be converted to RDF, as described
in~\cref{lpg_rdf_transform_section}. To benefit from different LPG features
while keeping RDF advantages such as simplicity, some researchers proposed and
implemented modifications to RDF. Examples are triple attributes or attaching
triples to other triples (described in~\cref{ab_section}).

Among native graph databases, while no LPG focused system simultaneously
supports RDF, some RDF systems (e.g., Amazon Neptune) also support LPG. 
Further, there has been recent work on unifying RDF and
LPG~\cite{lassila2023onegraph, gelling2023bridging, angles2022multilayer}.
Many other classes (KV stores, document stores, RDBMS, wide-column stores,
OODBMS) offer only LPG (with some exceptions, e.g., Oracle Spatial and Graph).
The latter suggests that it may be easier to express the LPG datasets than the
RDF datasets with the respective non-graph data models such as a collection of
documents.

Another interesting challenge is to understand better the design tradeoffs between
LPG and RDF. For example, it is unclear which one is more advantageous for different
workload classes, under different design constraints (disk vs.~in-memory representation,
distributed vs.~shared-memory, replicated vs.~sharded, etc.). This could be achieved
by developing formal runtime and storage models followed by an extensive evaluation.

There are very few systems that use neither RDF nor LPG.
HyperGraphDB uses the hypergraph model and GBase uses a simple directed graph
model without any labels or properties. Thus, these models are
of less relevance to practitioners, but they offer a potentially interesting
direction for researchers. Especially in the context of hypergraphs, there has
been a recent proliferation of novel schemes in other domains of graph
processing, such as graph
learning~\cite{baiHypergraphConvolutionHypergraph2021}, suggesting that
exploring hypergraphs for graph databases may be a timely direction.

\subsubsection{Graph Structure Representations}

Many graph databases use variants of
AL since it makes traversing neighborhoods efficient and
straightforward~\cite{neo4j_book}. This includes several (but not all) systems in the classes of LPG based native graph databases,
KV stores, document stores, wide-column stores, tuple stores, and OODBMS.
Contrarily, none of the considered RDF, RDBMS, and data hub systems explicitly use
AL. This is because the default design of the underlying data model, e.g., tables in RDBMS or documents
in document stores, do not often use AL.
%

\iftr
Moreover, none of the systems that we analyzed use an \emph{uncompressed} AM as it
is inefficient with $\mathcal{O}(n^2)$ space, especially for sparse graphs.
Systems using AM focus on compression of the adjacency
matrix~\cite{besta2018survey}, trying to mitigate storage and query overheads
(e.g., GBase~\cite{gbase_paper}).
\fi
%

\iftr
In AL, a potential cause for inefficiency is scanning all edges to find
neighbors of a given vertex. To alleviate this, index structures are
employed~\cite{besta2018log}. For a graph with $n$ vertices, such
an index is an array of pointers to respective neighborhoods, taking only
$O(n)$ space.
\fi

\subsection{Discussion, Takeaways, and Insights on Data Optimizations}
\label{discussion_opts}

We discuss separately common optimizations in data organization.

\subsubsection{Using Existing Storage Designs}

Most graph database systems are built upon existing storage designs, including
key-value stores, wide-column stores, RDBMS, and others. The advantage of
using existing storage designs is that these systems are usually mature and
well-tested. The disadvantage is that they may not be perfectly optimized for
graph data and graph queries. This is what native graph databases attempt to
address.
Overall, there is a lot of research potential in more efficient and more
effective use of existing storage designs for graph databases. For example, a
promising direction would be to investigate how recent RDBMS development, such
as worst-case optimal joins~\cite{ngo2018worst}, could be used for graph
related workloads.

\subsubsection{Data Layout of Records}

The details of record-based data organization heavily depend on a specific
system. For example, an RDBMS could treat a table row as a record, key-value
stores often maintain a single value in a single record, while in document
stores, a single document could be a record.
Importantly, some systems allow \textit{variable sized} records (e.g.,
ArangoDB), others only enable \textit{fixed sized} records (e.g., Neo4j).
Finally, we observe that while some systems (e.g., some triple stores such as
Cray Graph Engine) do not explicitly mention records, the data could still be
implicitly organized as records. In triple stores, one would
naturally associate a triple with a record.

Graph databases often use one or more records per vertex (these records are
sometimes referred to as \textit{vertex records}). Neo4j uses multiple
fixed-size records for vertices, while document databases use one document per vertex
(e.g., ArangoDB). Edges are sometimes stored in the same record together with
the associated (source or destination) vertices (e.g., Titan or JanusGraph).
Otherwise, edges are stored in separate \textit{edge records} (e.g., ArangoDB).

The records used by the studied graph databases may be unstructured (i.e.,
not having a pre-specified format such as JSON), as is the case with KV 
stores. They can also be structured: document databases often use the JSON
format, wide-column stores have a key-value mapping inside each row,
row-oriented RDBMS divide each row into columns, OODBMS impose some class
definition, and tuple stores as well as some RDF stores use tuples. 
The details of data layout (i.e., how vertices and edges are exactly
represented and encoded in records) may still vary across different system
classes.
Some structured systems still enable \emph{highly flexible} structure inside
their records. For example, document databases that use JSON or wide-columns
stores such as Titan and JanusGraph allow for different key-value mappings for
each vertex and edge. Other record based systems are more \emph{fixed} in
their structure. For example, in OODBMS, one has to define a class for each
configuration of vertex and edge properties. In RDBMS, one has to define tables
for each vertex or edge type.

Some systems (e.g., Sparksee, some triple stores, or column-oriented RDBMS)
do not store information about vertices and edges contiguously in dedicated
records. Instead, they maintain separate data structures for each property or
label. The information about a given vertex is thus distributed over different
structures. To find a property of a particular vertex, one has to query the
associated data structure (index) for that property and find the value for the
given vertex. Examples of such used index structures are B+ trees (in Sparksee)
or hashtables (in some RDF systems).

Overall, most systems use records to store vertices, most often one vertex
per one record. Some systems store edges in separate records, others store them
together with the adjacent vertices. To find a property of a particular vertex,
one has to find a record containing the vertex. The searched property is either
stored directly in that record, or its location is accessible via a pointer.
We observe that these design choices are made arbitrarily. We conclude that
an interesting research direction would be developing formal performance
models, and use them to guide the most advantageous design choice for each
type of storage backend for graph databases.

\subsubsection{Adjacencies between Records}

Another aspect of a graph data layout is the \emph{design of the adjacency
between records}. One can either assign each record an ID and then \emph{link
records to one another via IDs}, or one can \emph{use direct memory pointers}.
Using IDs requires an indexing structure to find the physical storage address
of a record associated with a particular ID. Direct memory pointers do not
require an index for a traversal from one record to its adjacent records.
%
%
Note that an index might still be used, for example to retrieve a vertex with a
particular property value (in this context, direct pointers only facilitate
resolving adjacency queries between vertices).

Using direct pointers can accelerate graph
traversals~\cite{neo4j_book}, as additional index traversals are
avoided.
Another option is to assign each record a unique ID and use these IDs instead
of direct pointers to refer to other records. On one hand, this requires an
additional indexing structure to find the physical location of a record based
on its ID. On the other hand, if the physical location changes, it is usually
easier to update the indexing structure instead of changing all associated
direct pointers, which may come with significant
overhead~\cite{arangodb_indexing_links}.
Here, one could also make these design choices and tradeoffs more accurate
by designing performance models to guide them.

\subsubsection{Storing Data Directly in Indexes}

Sometimes graph data is stored directly in an index. Triple stores use
indexes for various permutations of subject, predicate and object to answer
queries efficiently. Jena TBD stores its triple data inside of these indexes,
but has no triple table itself, since the indexes already store all necessary
data~\cite{apache_jena_tbd_links}. HyperGraphDB uses a key-value index,
namely Berkeley DB~\cite{berkeleydb}, to access its physical storage.
This approach also enables sharing primitive data values with a
reference count, so that multiple identical values are stored only
once~\cite{hypergraphdb_links}.


\subsubsection{Storing Strings}

Many systems, for example RDFs, \emph{heavily use strings},
for example for keeping IDs of different parts of graph datasets. Thus, there
are different schemes or maintaining strings. First, many systems support
both fixed-size and variable-size strings. This enables more performance: the
former (that encode, e.g., IDs) can be kept and accessed rapidly in dedicated
structures that assume certain specific sizes, while the latter (that encode,
e.g., arbitrary text items) often use separate dynamic structures that are
slower to access but offer more flexibility~\cite{whitedb_links,
neo4j_book}. Such dynamic stores can be both in-memory structures, but they
could even be separate files~\cite{neo4j_book}.

The considered systems offer other string related optimizations. For
example, CGE optimizes how it stores strings from its
triples/quads. Storing multiple long strings per triple/quad is 
inefficient, considering the fact that many triples/quads may share strings.
Therefore, CGE -- similarly to many other RDF systems -- maintains a
dictionary that maps strings to unique 48-bit integer identifiers (HURIs).  For
this, two distributed hashtables are used (one for mapping strings to HURIs and
one for mapping HURIs to strings). When loading, the strings are sorted and
then assigned to HURIs. This allows integer comparisons (equal, greater,
smaller, etc.) to be used instead of more expensive string comparisons.
This approach is shared by, e.g., tuple stores such as WhiteDB.

RDF systems also harness external dedicated libraries for more effective
string management. For example, the RDF String library~\cite{rdf-string}
facilitates constructing RDF strings. Specifically, it automatically generates
string representations of specified values, together with appropriate URI
prefixes.  This library can be used to even encode multi-line strings, or
conduct string interpolation with embedded expressions. Another example
is RDF String Turtle~\cite{rdf-string-json}, a package that enables
conversions between the string-based and RDF JSON representations of RDF.

\subsubsection{Data Distribution}
\label{sec:discussion-data-dist}

Almost all considered systems support a multi server mode and data
replication. Data sharding is also widely supported, but there are some systems
that do not offer this feature. We expect that, with growing dataset sizes,
data sharding will ultimately become as common as data replication. Still, it
is more complex to provide. We observe that, while sharding is as widely
supported on graph databases based on non-graph data models (e.g., document
stores) as data replication, there is a significant fraction of native graph
databases (both RDF and LPG based) that offer replication but \emph{not}
sharding. This indicates that non-graph backends are usually more mature in
designs and thus more attractive for industry purposes and for practitioners,
simultaneously implying research potential in the native graph database
designs.
We also observe that certain systems offer some form of tradeoff between
replication and sharding. Specifically, OrientDB offers a form of sharding, in
which not all collections of documents have to be copied on each server.
However, OrientDB does \emph{not} enable sharding of the collections
\emph{themselves} (i.e., distributing one collection across many servers). If
an individual collection grows large, it is the responsibility of the user to
partition the collection to avoid any additional overheads.
Thus, it would be interesting to research how to automatize sharding of single document
collections, or even single (large) documents.
Another such example is Neo4j which supports replication and provides
\emph{certain level} of support for sharding. Specifically, the user can
partition the graph and store each partition in a separate database, limiting
data redundancy.

Here, another interesting research opportunity would be to accelerate graph
database workloads such as OLAP by harnessing \emph{partial} data replication.
Specifically, many OLAP graph database workloads such as BFS can be expressed
with linear algebra operations~\cite{kepner2016mathematical}, and these
workloads could benefit from the partial replication of the adjacency matrix,
for example as done in 2.5D \& -3D matrix
multiplications~\cite{DBLP:journals/corr/SolomonikH15,
solomonik2014tradeoffs}.

\subsubsection{Indexes}
\label{sec:dis-indexes}

Most graph database systems use indexes.
Now, systems based on non-graph backends, for example RDBMS or document stores,
usually rely on existing indexing infrastructure present in such systems.
Native graph databases employ index structures for the neighborhoods of each
vertex, often in the form of direct pointers~\cite{neo4j_book}.

\textbf{Neighborhood indexes} are used mostly to speed up the access
of adjacency lists to accelerate traversal queries. JanusGraph calls
these indexes vertex-centric. They are constructed specifically for
vertices, so that incident edges can be filtered
efficiently to match the traversal conditions~\cite{titan_link}.
While JanusGraph allows multiple vertex-centric indexes per vertex, each
optimized for different conditions, which are then chosen by the query
optimizer, simpler solutions exist as well. LiveGraph uses a two-level
hierarchy, where the first level distinguishes edges by their label,
before pointing to the actual physical storage~\cite{livegraph}.
Graphflow indexes the neighbors of a vertex into forward and backward
adjacency lists, where each list is first partitioned by the edge label,
and secondly by the label of the neighbor
vertex~\cite{graphflow}. Another example is Sparksee, which uses
various different index structures to find the adjacent vertices and
properties of a vertex~\cite{sparksee_paper}.

An interesting research opportunity is to design indexes for richer
``higher-order'' structural information, beyond plain neighborhoods. Specifically,
a recent wave of pattern matching schemes~\cite{mhedhbi2021lsqb, gms}
indicates the importance of higher-order graph structure, such as triangles
that each vertex belongs to. Indexing such information would significantly
speed up queries related to clique mining, dense subgraph discovery,
clustering, and many others.
While some work has been done in this respect~\cite{sasaki2022language},
many designs could be proposed.

\textbf{Data indexes} concern data beyond the neighborhood information,
and they can be used to accelerate query plan generation and query execution. It is
possible for example to index all vertices that have a specific property
(value). They are usually employed to speed up Business Intelligence
workloads (details on workloads are in~\cref{sec:queries-workloads}). Many triple stores, for example
AllegroGraph~\cite{allegro_graph_links}, provide all six
permutations of subject (S), predicate (P), and object (O) as well as additional
aggregated indexes. However, to reduce associated costs, other approaches
exist as well: TripleBit uses just two permutations (PSO, POS) with two
aggregated indexes (SP, SO) and two auxiliary index
structures~\cite{triplebit_links}. gStore implements pattern
matching queries with the help of two index structures: a VS*-tree,
which is a specialized B+-tree, and a trie-based
T-index~\cite{gstore}. Some database systems like Amazon
Neptune~\cite{amazon_neptune_links} or
AnzoGraph~\cite{anzo_graph_links} only provide implicit indexes,
while still being confident to answer all kinds of queries efficiently.
However, most graph database systems allow the user to explicitly define
data indexes. Some of them, like Azure Cosmos
DB~\cite{azure_cosmosdb_links}, support composite indexes (a
combination of different labels/properties) for more specific use cases.
In addition to internal indexes, some systems employ external indexing
tools. For example, Titan and JanusGraph~\cite{titan_link} use
internal indexing for label- and value-based lookups, but rely on
external indexing backends (e.g.,
Elasticsearch~\cite{elasticsearch_links} or Apache
Solr~\cite{solr_links}) for non-trivial lookups involving
multiple properties, ranges, or full-text search.

\textbf{Structural indexes} are used for various internal data.
Here, LiveGraph uses a vertex index to map its vertex IDs to a
physical storage location~\cite{livegraph}. ArangoDB uses a hybrid
index, a hashtable, to find the documents of incident edges and adjacent
vertices of a vertex~\cite{arangodb_links}.

We categorize
systems (for which we were able to find this information)
according to this criteria in Table~\ref{tab:index_implementation}.
We find no clear connection between the index type and the backend of a
graph database, but most systems use tree based indexes.
A research opportunity, useful especially for practitioners,
would be to conduct a detailed and broad performance comparison
of different index implementations for different workloads.

\begin{table}[t]
\centering
\scriptsize
\sf
\begin{tabular}{lllll}
\toprule
\textbf{Graph Database System} & \textbf{Tree} & \textbf{Hashtable} & \textbf{Skip list} & \textbf{Additional remarks} \\
\midrule
Apache Jena TBD & \faThumbsUp$^*$ & \faTimes                 & \faTimes        & $^*$B+-tree \\
ArangoDB        & \faTimes        & \faThumbsUp$^*$          & \faThumbsUp$^*$ & $^*$depends on the used index engine \\
Blazegraph      & \faThumbsUp$^*$ & \faTimes                 & \faTimes        & $^*$B+-tree \\
Dgraph          & \faTimes        & \faThumbsUp              & \faTimes        & \\
Memgraph        & \faTimes        & \faTimes                 & \faThumbsUp     & \\
OrientDB        & \faThumbsUp$^*$ & \faThumbsUp$^{\ddagger}$ & \faTimes        & $^*$SB-tree with base 500 \\
                &                 &                          &                 & $^{\ddagger}$also supports a distributed hashtable index \\
VelocityGraph   & \faThumbsUp$^*$ & \faTimes                 & \faTimes        & $^*$B-tree \\
Virtuoso        & \faThumbsUp$^*$ & \faTimes                 & \faTimes        & $^*$2D R-tree \\
WhiteDB         & \faThumbsUp$^*$ & \faTimes                 & \faTimes        & $^*$T-tree \\
\bottomrule
\end{tabular}
\caption{
\textbf{Support for different index implementations in different graph database systems}.
``\faThumbsUp'': A system supports a given index implementation.
``\faTimes'': A system does not support a given index implementation.
}
\label{tab:index_implementation}
\end{table}

\subsubsection{Data Organization vs.~Database Performance}

Record based systems usually deliver more performance for queries that need to
retrieve all or most information about a vertex or an edge. They are more
efficient because the required data is stored in consecutive memory blocks. In
systems that store data in indexes, one queries a data structure per property, which
results in a more random access pattern. On the other hand, if one only wants
to retrieve single properties about vertices or edges, such
systems may only have to retrieve a single value. Contrarily, many record based
systems cannot retrieve only parts of records, fetching more data than
necessary.

Furthermore, a decision on whether to use IDs versus direct memory pointers to
link records depends on the read/write ratio of the workload for the given
system. In the former case, one has to use an indexing structure to find the
address of the record. This slows down read queries compared to following
direct pointers.  However, write queries can be more efficient with the use of
IDs instead of pointers. For example, when a record has to be moved to a new
address, all pointers to this record need to be updated to reflect this new
address. IDs could remain the same, only the indexing structure needs to modify
the address of the given record.

The available performance studies~\cite{Vojtech_masterthesis_comp_gdb,
vaikuntam2014evaluation, mccoll2014performance, ldbc, maschhoff2017quantifying} indicate that
systems based on non-graph data models, for example document
stores or wide-column stores, usually achieve more performance
for transactional workloads that update the graph. Contrarily,
read-only workloads (both simple and global analytics)
often achieve more performance on native graph stores.
Global analytics particularly benefit from native graph
stores that ensure parallelization of single queries~\cite{maschhoff2017quantifying}.
It underlies the potential and research opportunities for developing
hybrid database systems for graph workloads, that could combine the advantages
of databases using non-graph data models and of native graph stores.

\subsection{Discussion and Takeaways on Query Execution}
\label{sec:discussion-query-exec}

We discuss the query execution aspects of our taxonomy with respect
to the specific graph databases.
%
%
Our discussion is by necessity brief, as most systems do not disclose this
information\footnote{\scriptsize There is usually much more information
available on the data layout of a graph database, and \emph{not} its execution
engine.}.

\subsubsection{Concurrency \& Parallelization}

We start with concurrency and parallelization.

\textbf{Support for Concurrency and Parallelism in OLTP Queries}
We conclude that (1) almost all systems support concurrent OLTP
queries, and (2) in almost all classes of systems, fewer systems
support parallel OLTP query execution (with the exception of OODBMS based graph
databases). This indicates that more databases put more stress on high
throughput of queries executed per time unit rather than on lowering the
latency of a single query. A notable exception is the Cray Graph Engine, which does
\emph{not} support concurrent queries, but it \emph{does} offer parallelization
of single queries. In general, we expect most systems to ultimately support both
features.
With the growing importance of more complex OLTP workloads and the ongoing
process of blurring the difference between complex OLTP and Business Intelligence
workloads, putting more focus on parallel OLAP designs is becoming a more
attractive and urgent research direction~\cite{DBLP:journals/pvldb/SzarnyasWSSBWZB22}.



\textbf{Support for Concurrency and Parallelism in OLAP Queries}
OLAP queries are usually parallelized. This is because such queries often
involve all the vertices and edges, making the sequential execution
prohibitively long. Moreover, parallelization of such queries is facilitated
by the fact that there exists a plethora of related work, due to the prevalence
of certain OLAP queries (e.g., traversals, centralities) in static graph
analytics~\cite{kalavri2017high, batarfi2015large, mccune2015thinking,
kepner2016mathematical, besta2017push}.
At the same time, concurrent execution of different OLAP queries is supported 
only to some degree. For example, Weaver supports concurrent large-scale
OLAP queries such as BFS, but each such query processes an immutable separate
snapshot of the graph dataset.
One could attempt to investigate how to effectively run concurrent OLAP
queries (as well as large-scale read-only Business Intelligence workloads 
that have similar characteristics), which potentially only needs certain
amount of basic state bookkeeping.

\textbf{Support for Concurrent OLAP \& OLTP Queries}
%
%
Concurrent execution of OLAP and OLTP queries is not widely supported.
Systems, that support multi-versioning such as LiveGraph or Weaver, run
an OLAP query on a consistent graph snapshot, and current OLTP queries
modify different vertex and/or edge versions or introduce newer
versions. Other systems such as Neo4j, while allowing concurrent OLTP
and OLAP queries, leave dealing with inconsistencies of OLAP queries to
the client as the default isolation level (read-committed) does not
protect such queries from modification by other queries or even offers
repeatable reads.
This aspect is potentially rich in research and new design opportunities,
as it requires fundamental understanding of different effects that could occur
in native graph stores when executing concurrent OLAP and OLTP, analogously to
the effects happening at different isolation levels in RDBMS systems.

\textbf{Implementing Concurrent Execution}
One of the methods for query concurrency are different types of
locks. For example, WhiteDB provides database wide locking with a
\iftr
reader-writer lock~\cite{whitedb_links, rw_lock_book} 
\fi
which enables
concurrent readers but only one writer at a time.  As an alternative to locking
the whole database, one can also update fields of tuples atomically (set,
compare and set, add). WhiteDB itself does not enforce consistency, it is up to
the user to use locks and atomics correctly.
Another method is based on transactions, used for example by OrientDB that
provides distributed transactions with ACID semantics. We discuss transactions
separately in~\cref{sec:discussion-transact}.
Here, an interesting research opportunity would be to harness
lock-free synchronization protocols known from parallel
computing~\cite{blelloch2010parallel} when implementing different
fine-gained OLTP queries.

\textbf{Optimizing Parallel Execution}
Some of the systems that support parallel query execution explicitly
optimize the amount of data communicated when executing such parallelized
queries.
For example, the computation in CGE is distributed over
the participating processes. To minimize the amount of all-to-all
communication, query results are aggregated locally and -- whenever possible --
each process only communicates with a few peers to avoid network congestion.
%
%
Another way to minimize communication, used by MS Graph Engine and the
underlying Trinity database, is to reduce the sizes of data chunks exchanged by
processes.
For this, Trinity maintains special \emph{accessors} that allow for
accessing single attributes within a cell without needing to load the complete
cell. This lowers the I/O cost for many operations that do not need the whole
cells.
Several systems harness \emph{one-sided communication}, enabling processes
to access one another's data directly~\cite{gerstenberger2014enabling}.
For example, Trinity can be deployed on
InfiniBand~\cite{infiniband2015} to leverage Remote Direct Memory Access
(RDMA)~\cite{gerstenberger2014enabling}.
%
%
Similarly, Cray's infrastructure makes memory resources of
multiple compute nodes available as a single global address space, also
enabling one-sided communication in CGE. This facilitates parallel programming
\iftr
in a distributed environment~\cite{schmid2016high, gerstenberger2014enabling, besta2014fault}.
\fi

Here, a promising research opportunity is to harness established optimized
communication routines such as collectives~\cite{chan2007collective} for large-scale
OLAP and BI.

\textbf{Other Execution Optimizations}
The considered databases come with numerous other system-specific design
optimizations.
For example, an optimization in ArangoDB's design
allows to skip accessing the vertex document
and enables directly accessing one edge
document based on the vertex ID \emph{within another edge document}. This may
improve cache efficiency and thus reduce query execution
time~\cite{arangodb_indexing_links}.
Another example is Oracle Spatial and Graph that offers an interesting option
of \emph{switching its data backend based on the query being executed}.
Specifically, its in-memory analysis is boosted by the possibility to switch
the underlying relational storage with the native graph storage provided by the
PGX processing engine~\cite{hong2015pgx, roth2017pgx, el2019dsl}. In
such a configuration, Oracle Spatial and Graph effectively becomes a native
graph database. PGX comes with two variants, PGX.D and PGX.SM, that --
respectively -- offer distributed and shared-memory processing
capabilities~\cite{hong2015pgx}.
Furthermore, some systems use the LPG specifics to implement OLAP queries
more effectively. For example, to implement BFS, Weaver uses special designated
properties associated with vertices to indicate, whether a vertex has already
been already visited. Such special properties are not visible to an external
user of the graph database and are usually deallocated after a given query is
finalized.

A lot of work has been done into optimizing distributed-memory algebraic
operations such as matrix products using optimal communication
routines~\cite{Georganas:2012:CAO:2388996.2389132,
kwasniewski2021parallel, kwasniewski2021pebbles, kwasniewski2019red,
DBLP:journals/corr/SolomonikH15, solomonik2014tradeoffs, gleinig2022io}. It
would be interesting to investigate how such routines can be used to speed up
OLAP graph queries.

There is a large body of existing work in the design of dynamic graph
processing frameworks~\cite{besta2019practice}. These systems differ
from graph databases in several aspects, for example they often employ simple
graph models (and not LPG or RDF).
Simultaneously, they share the fundamental property of graph databases: dealing
with a dynamic graph with evolving structure. Moreover, different performance
analyses indicate that streaming frameworks are much faster (up to orders of
magnitude) than graph databases, especially in the context of raw graph updates
per second~\cite{vaikuntam2014evaluation,
mccoll2014performance}. This suggest that harnessing mechanisms used in such
frameworks in the context of graph databases could significantly enhance the
performance of the latter, and is a potentially fruitful research direction.

Furthermore, while there exists past research into the impact of the underlying
network on the performance of a distributed graph analytics
framework~\cite{ousterhout2015making}, little was done into investigating this
performance relationship in the context of graph database workloads. To the
best of our knowledge, there are no efforts into developing a topology-aware or
routing-aware data distribution scheme for graph databases, especially in the
context of recently proposed data center and high-performance computing network
topologies~\cite{besta2014slim, kim2008technology} and routing
architectures~\cite{besta2020highr}.

Finally, contrarily to the general static graph processing and graph
streaming, little research exists into accelerating graph databases using
different types of hardware architectures, accelerators, and hardware-related
designs, for example FPGAs~\cite{besta2019graph}, designs
related to network interface cards such as SmartNICs~\cite{di2019network}, or
processing in memory~\cite{ahn2016scalable}.

\subsubsection{ACID}
\label{sec:discussion-transact}

We also discuss various aspects of ACID transactions. ACID transactions
are usually used to implement OLTP queries. OLAP queries are read-only
analytics, and thus are less relevant for the notion of ACID. For example, 
in Weaver, OLAP workloads are referred to as ``node programs'' and are
treated differently from transactions, which are used to implement OLTP
queries.

\textbf{Support}
Overall, support for ACID transactions is widespread in graph databases.
However, there are some differences between respective system classes.  For
example, \emph{all} considered document and RDBMS graph databases offer full
ACID support. Contrarily, only around half of all considered key-value and
wide-column based systems support ACID transactions. This could be caused by
the fact that some backends have more mature transaction related designs.

\textbf{Implementation}
%
%
Two important ways to implement transactions are through locking or
timestamps. Neo4j uses write locks to protect modifications until they
are committed. Weaver uses timestamps to reorder transactions into an
serializable order if necessary. Other systems such as LiveGraph combine
both ways, and use timestamps to select the correct version of a
vertex/edge (solving simple read/write conflicts) and vertex locks to
deal with concurrent writes.

\subsubsection{Support for OLAP and OLTP Queries}
\label{sec:discussion-queries}

We analyze support for OLTP and OLAP. Both categories are widely supported,
but with certain differences across specific backend classes, specifically, (1)
all considered document stores focus solely on OLTP, (2) some RDBMS graph
databases do not support or focus on OLAP, and (3) some native graph databases
do not support OLTP. We conjecture that this is caused by the prevalent
historic use cases of these systems, and the associated features of the
backend design. For example, document stores have traditionally mostly
focused on maintaining document related data and to answer simple queries,
instead of running complicated global graph analytics. Thus, it may be very
challenging to ensure high performance of such global workloads on this
backend class. Instead, native graph databases work directly with the graph
data model, making it simpler to develop fast traversals and other OLAP
workloads.
As for RDBMS, they were traditionally not associated with graph global
workloads. However, graph analytics based on RDBMS
has become a separate and growing area of research. Zhao et
al.~\cite{rdbms_allinone_paper} study the general use of RDBMS for
graphs. They define four new relational algebra operations for modeling graph
operations. They show how to define these four operations with six smaller
building blocks: basic relational algebra operations, such as group-by and
aggregation. Xirogiannopoulos et al.~\cite{graphgen_rdbms_paper}
describe GraphGen, an end-to-end graph analysis framework that is built on top
of an RDBMS. GraphGen supports graph queries through so called Graph-Views that
define graphs as transformations over underlying relational datasets. This
provides a graph modeling abstraction, and the underlying representation can be
optimized independently.

Some document stores still provide at least partial support for traversal-like
workloads. For example, in ArangoDB, documents are indexed using a hashtable, where the
\textit{\_key} attribute serves as the hashtable key.
A traversal over the neighbors of a given vertex works as follows. First, given
the \textit{\_key} of a vertex~$v$, ArangoDB finds all $v$'s adjacent edges
using the hybrid index. Next, the system retrieves the corresponding edge
documents and fetches all the associated \textit{\_to} properties. Finally, the
\textit{\_to} properties serve as the new \textit{\_key} properties when
searching for the neighboring vertices.
%

There are other correlations between supported workloads and system design
features. For instance, we observe that systems that do not target OLTP, also
often do not provide, or focus on, ACID transactions. This is because ACID is
not commonly used with OLAP. Examples include Cray Graph Engine, RedisGraph,
or Graphflow.

There also exist many OLAP graph workloads that have been largely unaddressed by
the design and performance analyses of existing graph database systems. 
This includes
vertex reordering problems (e.g., listing vertices by their
degeneracy), or optimization (e.g., graph
coloring)~\cite{besta2020highcolor}.
There problems were considered in the context of graph algorithms processing
simple graphs, and incorporating rich models such as RDF would further
increase complexity, and offer many associated research challenges,
for example designing indexes, data layouts, or distribution
strategies.

\subsubsection{Supported Languages}
\label{sec:discussion-languages}

We also analyze support for graph query languages.  Some types of backends
focus on one specific language: triple stores and SPARQL, document stores and
Gremlin, wide-column stores and Gremlin, RDBMS and SQL.  Other classes are not
distinctively correlated with some specific language, although Cypher seems
most popular among LPG based native graph stores.
Usually, the query language support is primarily affected by the supported
conceptual graph model; if it is RDF, then the system usually supports SPARQL
while systems focusing on LPG often support Cypher or Gremlin.

Several systems come with their own languages, or variants of the
established ones. For example, in MS Graph Engine, cells are associated with a
schema that is defined using the Trinity Specification Language
(TSL)~\cite{trinity_paper}. TSL enables defining the structure of cells
similarly to C-structs. For example, a cell can hold data items of different
data types, including IDs of other cells.
Moreover, querying graphs in Oracle Spatial and Graph is possible using
PGQL~\cite{van2016pgql}, a declarative, SQL-like, graph pattern matching
query language. PGQL is designed to match the hybrid structure of Oracle
Spatial and Graph, and it allows for querying both data stored on disk in
Oracle Database as well as in in-memory parts of graph datasets.

Besides their primary language, systems also offer support for additional
language functionalities. For example, Oracle Spatial and Graph also supports
SQL and SPARQL (for RDF graphs). Moreover, the offered Java API implements
Apache Tinkerpop interfaces, including the Gremlin API.

\subsection{Insights for Practitioners}

An important question on whether RDBMS or non-relational native graph
backends are more suitable for graph workloads, is far from being fully
answered.
On one hand, several analyses~\cite{pacaci2017we,
aberger2017emptyheaded, jindal2015graph, ibm_db2_graph, fan2015case} --
including very recent ones~\cite{ten2023duckpgq} -- indicate better
performance of RDBMS over native graph database designs.
However, these analyses focus on systems designed with a single class of
workloads in mind (e.g., OLAP analytics), and on homogeneous graphs without
rich additional label and property data.
Thus, they are not conclusive for more realistic scenarions where a mix of
OLAP, OLTP, and BI workloads runs over rich LPG datasets.
This is supported by another recent study, which states that \emph{``The
workloads (...) require several storage and processing features that existing
RDBMSs are generally not optimized for.''}~\cite{fengkuzu}.

In general, whenever the data schema is known in advance, RDBMS -- with its
long-standing history of optimizations for such cases -- would be a
preferable choice.
Contrarily, when the data schema is not known, graph databases would most
probably offer more performance.

Overall, systems based on non-graph data models, such as RDBMS or -- to a
certain degree -- others (e.g., document stores) offer most mature designs and
well-understood behavior related to different isolation levels. As such, these
systems are the best option when one needs a system that offers predictable
behavior in the first place, and reasonable performance for standard
workloads.
On the other hand, when one aims at highest performance of purely graph
workloads, it is worth considering native graph stores. This is especially
the case for the most recent graph workload classes that are only now being
introduced in the GDB landscape, such as subgraph queries~\cite{mhedhbi2021lsqb}.

\section{CONCLUSION}

Graph databases constitute an important area of academic research and different
industry efforts. They are used to maintain, query, and analyze numerous
datasets in different domains in industry and academia. Many graph
databases of different types have been developed. They use many data models and
representations, they are constructed using miscellaneous design choices, and
they enable a large number of queries and workloads. In this work, we provide the first survey
and taxonomy of this rich graph database landscape. Our work can be
used not only by researchers willing to learn more about this fascinating subject, but
also by architects, developers, and project managers who want to select the
most advantageous graph database system or design.

\iftr

%
\begin{acks}
We thank Gabor Szarnyas for extensive feedback, and Hannes Voigt, Matteo
Lissandrini, Daniel Ritter, Lukas Braun, Janez Ales, Nikolay Yakovets, and
Khuzaima Daudjee for insightful remarks.
We thank Timo Schneider for immense help with infrastructure at SPCL,
and PRODYNA AG (Darko Križić, Jens Nixdorf, Christoph Körner) for generous
support.
This project received funding from the European Research Council
\raisebox{-0.25em}{\includegraphics[height=1em]{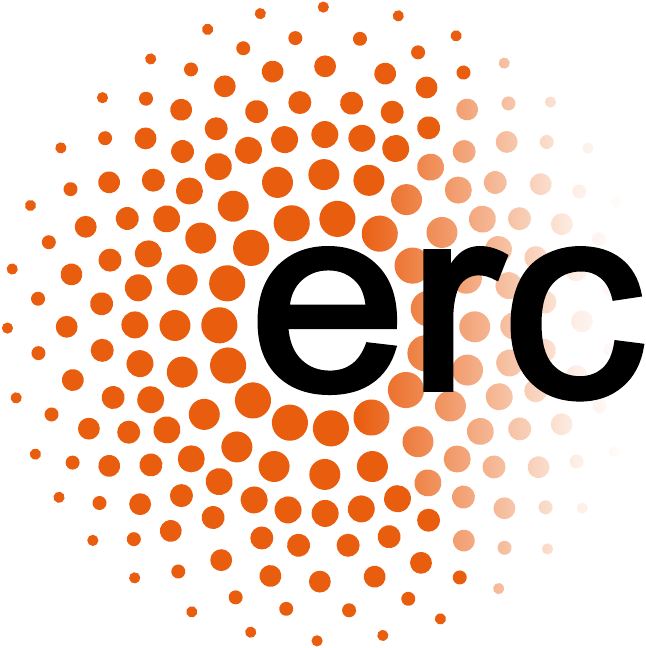}} (Project PSAP,
No.~101002047), and the European High-Performance Computing Joint Undertaking
(JU) under grant agreement No.~955513 (MAELSTROM).
This project was supported by the ETH Future Computing Laboratory (EFCL),
financed by a donation from Huawei Technologies.
This project received funding from the European Union's HE research and
innovation programme under the grant agreement No.~101070141 (Project
GLACIATION).
\end{acks}

\fi

\iftr

{
\bibliographystyle{ACM-Reference-Format}
\bibliography{references.complete}
}

\fi

\maciej{ADD: The Current State of Graph Databases. See below and emails
reification rdf, google and add.
}

\maciej{REIFICATION mention}

\maciej{check comments from Lukas, and this guy that sent email}

\maciej{Go over this Neo4j Jesus article and consider all these differences
between RDF and LPG?}

\maciej{Survey of Graph Database Performance on the HPC Scalable Graph Analysis Benchmark
Authors
Authors and affiliations
D. Dominguez-SalP. Urbón-BayesA. Giménez-VañóS. Gómez-VillamorN. Martínez-BazánJ. L. Larriba-Pey}

\maciej{TODO: discuss "multi-model"
https://orientdb.com/graph-database/}

\maciej{Check: A Performance Evaluation of Open Source Graph Databases,
A Brief Study of Open Source Graph Databases, An introduction to Graph Data Management
Efficient
data structures for dynamic graph
analysis
}

\maciej{NEW PAPERS:
Hermes: Dynamic Partitioning for Distributed Social Network Graph Databases.
The Current State of Graph Databases Mike Buerli
Towards scalable distributed graph database engine for hybrid clouds
Transaction Management for Cloud-Based Graph Databases
Database Research Challenges and Opportunities of Big Graph Data
Comparative Analysis of Relational and Graph Databases for Social Networks
DistNeo4j: Scaling Graph Databases through Dynamic Distributed Partitioning
LiveGraph: A Transactional Graph Storage System with Purely Sequential Adjacency List Scans
Views and Transactional Storage for Large Graphs
IOGP: An incremental online graph partitioning algorithm for distributed graph databases
Cascade-aware partitioning of large graph databases
}

\maciej{TODO maybe in the future (or just describe in related work):
Systems mentioned in various listings and
rankings~\cite{pat_links,dbguide_links,dreamcss_links,g2crowd_links,dbengines_links},
but without active websites or reports:
MapGraph,
towards a distributed large scale dynamic graph data store (DegAwareRHH),
HOW DO I CHOOSE THE RIGHT NOSQL SOLUTION? A
COMPREHENSIVE THEORETICAL AND EXPERIMENTAL
SURVEY,
Using Domain-Specific Languages For Analytic Graph
Databases,
System G Distributed Graph Database,
Sikos: Graph Databases, Chapter 6,
Filament (relational DBMS as backend),
IBM System G Native Store (property and RDF graphs),
grapholytic (object-oriented),
k-infinity,
SPARQLVerse,
IBM Graph~\cite{ibm_graph_links} (retired 2017)
GlobalsDB~\cite{globalsdb_links} (GitHub repository only),
Objectivity InfiniteGraph~\cite{infinite_graph_links} (precurser to Objectivity ThingSpan).}

\maciej{TODO: from nono's comments:
Btw. żeby mi nie umkneło. W surveyu o NoSQL jest takie coś:
"In most graph stores, data partitioning either is not included or simply
implemented through random partitioning."
Jeśli to jest faktycznie słabo ogarnięte, to może też być ciekawy temat. Samo
partycjonowanie grafu wiadomo, NP-zupełne i dobrze opracowane. Ale w kontekście
workloadów na rzeczywistych bazach to może być inna bajka. W sensie, że może da
się jakoś punktować edge, które często wspólnie występują w wyszukiwaniach
(albo często występują w wyszukiwaniach związanych z jakimś nodem), i przenosić
je na wspólne partitiony.
Swoją drogą, to też mogłoby łapać się w kontekst tego, jak sposób
przechowywania graph data wpływa na szybkość danych kategorii wyszukań. Trzeba
to szybko zrobić, bo to fest ciekawe.
}

\end{document}